\documentclass[10pt,twocolumn,amsmath,amssymb,aps]{revtex4}

\usepackage{graphicx}       
\usepackage{dcolumn}        
\usepackage{bm}             
\usepackage{psfrag}
\usepackage{tensor}

\newcommand{\nn}{\nonumber}

\newcommand{\gam}{\gamma}
\newcommand{\bet}{\beta}
\newcommand{\alp}{\alpha}

\newcommand{\beq}{\begin{equation}}
\newcommand{\eeq}{\end{equation}}

\newcommand{\Lag}{\mathcal{L}}

\newcommand{\FE}{\mathcal{F}_{(E)}}
\newcommand{\FJ}{\mathcal{F}_{(J)}}
\newcommand{\J}{J_z}
\newcommand{\CE}{C_{(E)}}
\newcommand{\CJ}{C_{(J)}}

\newcommand{\pml}{\gamma}
\newcommand{\xmid}{x_0}
\newcommand{\xwid}{\sigma_x}
\newcommand{\xinner}{x_{+}}

\begin{document}

\preprint{}

 \title{Superradiant instabilities of rotating black holes in the time domain}

\author{Sam R. Dolan}
 \email{s.dolan@shef.ac.uk}
 \affiliation{%
Consortium for Fundamental Physics, School of Mathematics and Statistics, 
University of Sheffield, Hicks Building, Hounsfield Road, Sheffield S3 7RH, United Kingdom 
  \\
}%

\date{\today}

\begin{abstract}
Bosonic fields on rotating black hole spacetimes are subject to amplification by superradiance, which induces exponentially-growing instabilities (the `black hole bomb') in two scenarios: if the black hole is enclosed by a mirror, or if the bosonic field has rest mass. Here we present a time-domain study of the scalar field on Kerr spacetime which probes ultra-long timescales up to $t \lesssim 5 \times 10^6 M$, to reveal the growth of the instability. We describe an highly-efficient method for evolving the field, based on a spectral decomposition into a coupled set of 1+1D equations, and an absorbing boundary condition inspired by the `perfectly-matched layers' paradigm. First, we examine the mirror case to study how the instability timescale and mode structure depend on mirror radius. Next, we examine the massive-field, whose rich spectrum (revealed through Fourier analysis) generates `beating' effects which disguise the instability. We show that the instability is clearly revealed by tracking the stress-energy of the field in the exterior spacetime. We calculate the growth rate for a range of mass couplings, by applying a frequency-filter to isolate individual modal contributions to the time-domain signal. Our results are in accord with previous frequency-domain studies which put the maximum growth rate at $\tau^{-1} \approx 1.72 \times 10^{-7} (GM/c^3)^{-1}$ for the massive scalar field on Kerr spacetime. 
\end{abstract}

\maketitle

%
%

\section{\label{sec:introduction}Introduction}
An exciting era is underway in astrophysics, in which precision measurements of black hole masses and spins are becoming possible \cite{Shafee, Miller, Brenneman, Walton}. Naturally, there is renewed interest in the range of physical processes that may occur in the vicinity of black holes.
One of the most intriguing processes is the so-called `black hole bomb' effect, first proposed by Press and Teukolsky in 1972 \cite{Press:1972}, in which the rotational energy of a Kerr black hole (BH) \cite{Kerr:1963} generates runaway amplification in a bosonic (i.e.~integer-spin) field. The driving mechanism is superradiance \cite{Starobinskii:1973}: the stimulated emission of low-frequency radiation in co-rotating modes. A perturbation of frequency $\omega$ and azimuthal number $m$, 
incident upon a black hole of mass $M$ and angular momentum $J = aM$ has a reflection coefficient of \emph{greater than unity} if the superradiance condition is met:
\beq
\omega \tilde{\omega} < 0, \quad \quad \label{eq:superrad}
\eeq
where $\tilde{\omega} \equiv \omega - m \Omega$ and $\Omega = a / (2 M r_+)$ is the angular frequency of the event horizon, with $r_+$ its radius.
Via superradiance, a black hole may shed mass and angular momentum, but in such a way that the horizon area ($A = 8 \pi M r_+$) actually \emph{increases}. Since horizon area is associated with entropy \cite{Bardeen}, superradiance may be thought of as a thermodynamically-favoured relaxation process. 

The Kerr spacetime in vacuum is (apparently) stable under perturbation by massless fields (see e.g.~\cite{Regge:1957, Press:1973, Whiting, Dafermos1}). For an instability to develop, some additional mechanism is needed to confine the field in the vicinity of the black hole. If the BH is surrounded by a reflecting boundary (i.e.~a `mirror'), then superradiant flux is reflected back on to the hole, generating an exponentially-growing instability \cite{Press:1972, Cardoso:2004-bomb}. A boundary to the spacetime, as in Anti-de Sitter scenarios, may play a similar role \cite{Cardoso:2004-AdS}. A more `natural' astrophysical scenario arises if the bosonic field has a non-zero rest mass $\mu$, in which case \emph{quasi-bound states} \cite{Deruelle:1974, Damour:1976, Zouros:1979, Detweiler:1980, Gaina:1992, Gaina:1993, Cardoso:Yoshida:2005, Lasenby:2005-bs, Dolan:2007, Grain:2007, Barranco} may form around the BH, with $\text{Re}(\omega) \lesssim \mu$.
Quasi-bound states are localized modes which are associated with a minimum in the effective potential, induced by the gravitational attraction between the field and BH mass \cite{Damour:1976}. 
The spectrum of quasi-bound states on the Kerr spacetime depends on just two dimensionless quantities: the BH spin $a / M$ and the mass coupling,
\beq
\frac{G M \mu}{\hbar c} \equiv M \mu,
\eeq
where in the following we adopt units $G = c = \hbar = 1$. 

There has recently been fresh interest in a fundamental question: is the superradiant instability in the massive bosonic field of \emph{any} relevance to astrophysics or particle theory? Here, we must be mindful of a strong constraint: the instability only applies if $M \mu \lesssim m M \Omega$, with $M \Omega \le 1/2$, and furthermore the growth rate is exponentially suppressed at large $m$ \cite{Zouros:1979}. In other words, the instability is only physically relevant if there is a bosonic field with a Compton wavelength comparable to (or larger than) the horizon scale. Writing $M \mu = 7.5 \times 10^9 (M / M_\odot) (\mu / [1 \text{eV}] )$, where $M_\odot$ is the solar mass, makes it clear that this condition is not satisfied by any known massive bosons in astrophysical black hole scenarios. However, there are two more exotic scenarios which may be relevant: standard-model massive bosons interacting with `primordial' black holes, or as-yet-undiscovered ultra-light bosons interacting with solar-mass or supermassive black holes. The latter scenario is motivated by compactifications of string theory, which suggest the existence of \emph{axions}: a family of ultra-light bosons with masses spanning across many decades, possibly all the way down to the Hubble scale $\mu \sim 10^{-33} \text{eV}$. If extant, axions could ignite the superradiant instability in stellar-mass or supermassive black holes, generating `gaps' in the black hole mass-spin spectrum \cite{Arvanitaki1, Arvanitaki2}. It has been suggested that precision measurement of the masses and spins of supermassive black holes can be used to put new constraints on the existence of ultra-light bosons \cite{Arvanitaki1, Arvanitaki2, Kodama:Yoshino, Pani:PRL, Pani:PRD}. It is argued that unexpected gaps in the BH spectrum would be  long-sought-after observational evidence in favour of string theory (or at least, in favour of a beyond-Standard-Model theory requiring compactification of higher dimensions which leads to a family of ultra-light bosons). 

The majority of the literature to date has focussed on the instability caused by the \emph{scalar} (i.e.~spinless) field \cite{Zouros:1979, Detweiler:1980, Furuhashi:2004, Cardoso:Yoshida:2005, Dolan:2007, Rosa:2010, Lee:2011, Li:2012, Arvanitaki1, Arvanitaki2, Kodama:Yoshino, Yoshino:Kodama}. The massive scalar-field equations on Kerr spacetime are separable in the frequency domain \cite{Brill:1972}, hence a rather complete picture of the instability in the linear regime may be obtained via straightforward analysis of \emph{ordinary} differential equations (ODEs).  In Sec.~\ref{subsec:bound-states} we review the frequency-domain formalism and the key results in some detail.

The aim of this work is to study the instability in the \emph{time domain}. Time domain methods offer a general toolkit for studying systems of partial differential equations of hyperbolic character, which are well-suited to cases where governing equations are non-separable and/or non-linear.

An interesting example of an apparently \emph{non-separable} but linear system is the massive vector (Proca) field on Kerr spacetime. 
In the Proca-Schwarzschild case, the equations may be separated, by decomposing the field in vector spherical harmonics, leading to a single ODE describing the odd-parity part and a pair of coupled ODEs for the even-parity part \cite{Rosa:Dolan, Herdeiro}. In the Proca-Kerr case, impressive recent progress \cite{Pani:PRL,Pani:PRD} has been made by expanding the equations in powers of $a/M$, applying a method originally developed to analyze radiation reaction in rotating neutron stars \cite{Kojima:1992}. In a technical tour-de-force \cite{Pani:PRL, Pani:PRD}, it was shown that, at first and second orders in $a/M$, the equations may once more be separated with vector harmonics. This has opened a window on to Proca-Kerr phenomenonology, which has led to a new constraint on the mass of the photon \cite{Pani:PRL} from measurements of supermassive BHs. New time-domain methods \cite{Witek} offer a way to move beyond the slow-rotation regime, and to test the applicability of the approximation.

\emph{Non-linear} effects will become important as the instability develops. An interesting question is: does the instability end in an explosive manner (with a `bosenova' \cite{Arvanitaki1}), or does the system relax towards a quasi-stable state? A recent time-domain study \cite{Yoshino:Kodama} examined the evolution of an axion field with a non-linear self-interaction, and concluded that the system is prone to a sudden collapse of the axion cloud with a rapid release of energy.

However, even within the linear regime, studying the onset of the scalar-field instability in the time domain is a difficult challenge, because (as anticipated from frequency-domain analyses \cite{Furuhashi:2004, Cardoso:Yoshida:2005, Dolan:2007, Rosa:2010}) the growth rate of the instability is rather slow, with a minimum e-folding time of $\sim 5.8 \times 10^6 M$ (see Sec.~\ref{subsec:bound-states} and Fig.~\ref{fig:freq-domain-growth}). To date, this timescale has proven to be beyond the reach of multi-dimensional simulations \cite{Strafuss:2005, Witek, Yoshino:Kodama}, which (so far) have evolved the field up to $t \lesssim 10^4M$ or less. Despite this limitation, a very recent time domain study \cite{Witek} seems to have revealed the onset of the Proca-Kerr instability, confirming that it develops much more rapidly than the scalar-field instability. A key aim of this work is to construct a time-domain scheme capable of probing ultra-long timescales of the order of $10^6 M$ or more. We will show that ultra-long evolutions, in combination with Fourier analysis, allow us to determine the physically-relevant part of the quasi-bound state spectrum to high accuracy. We will argue that the methods developed here are ripe for application in the Proca-Kerr case (amongst others).

Superradiant instabilities appear in a range of guises, for example, for BHs immersed in a magnetic field \cite{Konoplya:2009}, for higher-dimensional black holes, branes and strings \cite{Cardoso:2005-new, Cardoso:Yoshida:2005, Lee:2011, Rosa:2012}, black holes in Anti-de Sitter (AdS) spacetimes \cite{Cardoso:2004-AdS, Aliev:Delice:2008, Li:2012, Witek:box} and alternative spacetimes \cite{Konoplya:Zhidenko:2011} and alternative gravitational theories \cite{Myung:2011}. The repulsive effect of a superradiant instability on orbits around a black hole has also been considered \cite{Press:1972, Cardoso:Chakrabarti:2011}. 
Charged BHs generate superradiance in charged bosonic fields which leads to an enhanced superradiant instability on charged rotating spacetimes \cite{Damour:1976, Furuhashi:2004}. On the other hand, non-rotating charged black holes appear to be stable under this effect \cite{Hod:2012}. 
Given the range of possible scenarios, it is possible that the time-domain method developed here could have many applications. 

In Sec.~\ref{sec:foundations} we describe the basic setup and review the properties of quasi-bound states as revealed by frequency domain analyses. In Sec.~\ref{sec:methods} we develop a time-domain method for evolving the massive scalar field on Kerr. We give a selection of numerical results in Sec.~\ref{sec:results} and conclude with a discussion in Sec.~\ref{sec:conclusion}.


\section{Foundations\label{sec:foundations}}
In this section we recap the essential features of the scalar wave equation on the Kerr spacetime (\ref{subsec:waveeq}), and review the properties of the quasi-bound state spectrum in the frequency domain (\ref{subsec:bound-states}). 
\subsection{Spacetime, wave equation and separability\label{subsec:waveeq}}
The Kerr spacetime may be described via the Boyer-Lindquist coordinate system $\{ t , r, \theta, \phi \}$, in which the line element $ds^2 = g_{\alp \bet} dx^\alp dx^\bet$ is
\begin{eqnarray}
ds^2 = - \left(1 - \frac{2Mr}{\rho^2} \right) dt^2 - \frac{4 a M r \sin^2 \theta}{\rho^2} dt d\phi + \frac{\rho^2}{\Delta} dr^2 \nn \\
+ \rho^2 d\theta^2 + \left( r^2 + a^2 + \frac{2M r a^2 \sin^2 \theta}{\rho^2} \right) \sin^2 \theta d\phi^2 , \nn
\end{eqnarray}
where $\rho^2 = r^2 + a^2 \cos^2 \theta$ and $\Delta = r^2 - 2Mr + a^2$. The Klein-Gordon equation governing a scalar field $\Phi$ of mass $\mu$ is
\beq
\Box \Phi - \mu^2 \Phi = 0,  \label{eq:kg}
\eeq
where 
\beq
\Box \Phi \equiv \tensor{\Phi}{_{;\alp} ^{;\alp}} = \left(-g\right)^{-1/2} \left( (-g)^{1/2} g^{\alp \bet} \Phi_{, \alp}  \right)_{,\bet} ,
\eeq
and the subscripts `$;$' and `$,$' denote covariant and partial differentiation, respectively. Here $g = -\rho^4 \sin^2 \theta$ is the metric determinant, and $g^{\alp \bet}$ is the contravariant version of the metric.

In the frequency domain, Eq.~(\ref{eq:kg}) admits separable solutions \cite{Brill:1972, Teukolsky:1972} of the form 
\beq
\Phi(t,r,\theta,\phi) =  e^{-i \omega t} e^{i m \phi} R(r) S(\theta) ,
\eeq
where the radial and angular functions satisfy
\begin{eqnarray}
&& \frac{d}{dr} \left( \Delta \frac{dR}{dr} \right) + \left( K^2 / \Delta - \lambda - r^2 \mu^2 \right) R = 0 , \\
&& \frac{1}{\sin \theta} \frac{d}{d\theta} \left[ \sin \theta \frac{dS}{d\theta} \right] + \nn \\
&& \quad \left[ \Lambda + a^2 (\omega^2 - \mu^2) \cos^2 \theta - \frac{m^2}{\sin^2\theta} \right] S = 0 ,
\end{eqnarray}
with $K = (r^2+a^2) \omega - am$ and $\lambda = \Lambda - 2 a m \omega + a^2 \omega^2$.  The solutions of the angular equation, $S = S_{lm}(\theta)$, are spheroidal harmonics, and the angular eigenvalues can be found from a series expansion in the small-$a \omega$, $a\mu$ regime \cite{Seidel:1989} (with $\Lambda \rightarrow l(l+1)$ as $a\omega, a\mu \rightarrow 0$), or by solving a continued-fraction equation \cite{Leaver:1985}.

The radial equation can be written in standard 2nd-order form, by applying a simple transformation \cite{Teukolsky:1973, Sasaki:Nakamura}, $R = X / \sqrt{r^2 + a^2}$, to obtain
\beq
\frac{d^2 X}{dx^2} + U(x) X = 0, 
\eeq
with
\beq
U(x) = (r^2+a^2)^{-2} \left[ K^2 - \Delta (\lambda + \mu^2 r^2) \right] + \beta^2 - \partial_x \beta, 
\eeq
 where $\beta = r \Delta / (r^2 + a^2)^2$. Here, the tortoise coordinate 
 $x$ (often denoted $r_\ast$) is defined via
\beq
\frac{d x}{d r} = \frac{r^2 + a^2}{\Delta} ,
\eeq
or, after fixing the constant of integration,
\beq
x = r + \frac{2M}{r_+ - r_-} \left(r_+ \ln \left|\frac{r-r_+}{2M} \right| - r_- \ln \left| \frac{r - r_-}{2M} \right| \right) .
\eeq
In the near-horizon limit, $r \rightarrow r_+$, the physical solution satisfies the ingoing boundary condition,
\beq
X \sim \exp(- i \tilde{\omega} x) ,  \label{eq:ingoing}
\eeq
where $\tilde{\omega}$ was defined below Eq.~(\ref{eq:superrad}). In the far-field, the solution may be written as a linear superposition
\beq
X \sim A^+ \exp( q x) + A^- \exp(-q x) ,
\eeq
where $q = \sqrt{\mu^2 - \omega^2}$ and we define the root so that $\text{Re}( q ) \ge 0$. 

\subsection{Quasi-bound state spectrum\label{subsec:bound-states}}
The quasi-bound states are defined as the convergent solutions:  those for which $A^+_{lm} = 0$. The black hole is an open system as flux may pass into the horizon; hence it is natural that the quasi-bound states (like the quasi-normal modes \cite{Leaver:1985}) have \emph{complex} frequencies, 
\beq
\omega = \hat{\omega} + i \nu .
\eeq
Here, the real part $\hat{\omega}$ sets the oscillation frequency of the mode, and the imaginary part $\nu$ sets the growth ($\nu > 0$) or decay ($\nu < 0$) rate. The modes in the spectrum are labelled by integers for orbital angular momentum $l$, azimuthal angular momentum $m$, overtone number $n$ and, in the non-zero spin case, polarization $S$. If the oscillation frequency $\hat{\omega}$ lies \emph{inside} the superradiant regime ($\hat{\omega} \hat{\tilde{\omega}} < 0$) then the bound state will grow ($\nu > 0$); if $\hat{\omega} = m\Omega$ then the mode is stationary ($\nu = 0$); otherwise the mode will decay ($\nu < 0$).   All quasi-bound states on the non-rotating (Schwarzschild) spacetime are found to be decaying. We will confirm this picture in later sections.

The physically-relevant states grow ($\nu > 0$) or decay ($\nu < 0$) only slowly in comparison to the Compton and BH light-crossing times, i.e.~$M |\nu|, |\nu| / \mu \ll 1$.  Foundational work by Detweiler \cite{Detweiler:1980} and Zouros \& Eardley \cite{Zouros:1979} led to simple approximations for the growth rate in the low- and high-$M\mu$ regimes. These studies imply a maximum growth rate $\nu_{\text{max}}$ of approximately
\beq
M \nu_{\text{max}} \sim \begin{cases}
 \zeta (M\mu)^9 \left( a / M - 2 \mu r_+ \right),  & M \mu \ll 1 , \\
 10^{-7} e^{-1.84 M \mu} , & M \mu \gg 1 .
\end{cases}
\eeq
where $\zeta$ is a numerical constant \footnote{Ref.~\cite{Detweiler:1980} obtained $\zeta = 1/24$ (Eq.~(28), whereas other analyses \cite{Rosa:Dolan, Pani:PRD} suggest $\zeta = 1/48$.}.
The maximum growth rate occurs in the intermediate regime $M \mu \sim 1/2$, which has been explored with  numerical analyses. It turns out that, in close analogue to the quasinormal modes \cite{Leaver:1985, Simone:1992, Konoplya:2006}, the quasi-bound states are minimal solutions of a three-term recurrence relation \cite{Dolan:2007}. Finding minimal solutions is equivalent to finding the roots of a continued fraction equation; hence finding the spectrum numerically is `easy' \cite{Dolan:2007}. A range of studies, by Furuhashi \& Nambu \cite{Furuhashi:2004}, Cardoso \& Yoshida \cite{Cardoso:Yoshida:2005}, Dolan \cite{Dolan:2007}, Rosa \cite{Rosa:2010} and Kodama \& Yoshino \cite{Kodama:Yoshino}, have presented numerical results which are in accord: the maximum growth rate $\nu_{\text{max}}$ occurs in the fundamental mode $n=0$ of the co-rotating dipole mode $l=m=1$ for $\mu$ near to $\left| \Omega \right|$. The rate $\nu_{\text{max}}$ is strongly dependent on the black hole spin $a/M$.

Plots of the dependence of the growth rate $\nu$ on $a/M$ and $M\mu$ were previously been presented in (e.g.)~Fig.~6 in Ref.~\cite{Dolan:2007}, Fig.~2 in Ref.~\cite{Cardoso:Yoshida:2005}, Fig.~6 in Ref.~\cite{Furuhashi:2004} and Fig.~14 in Ref.~\cite{Kodama:Yoshino}. Typical values of the maximum growth rate for a range of spin rates $a < M$ were given in Table I of Ref.~\cite{Dolan:2007}. For example, the maximum rate at (e.g.) $a = 0.99M$ is $M \nu \approx 1.5 \times 10^{-7}$, occurring at $M \mu \approx 0.42$. The extremal case $a = M$ was considered in \cite{Rosa:2010}. 

\begin{figure}
 \includegraphics[width=8.2cm]{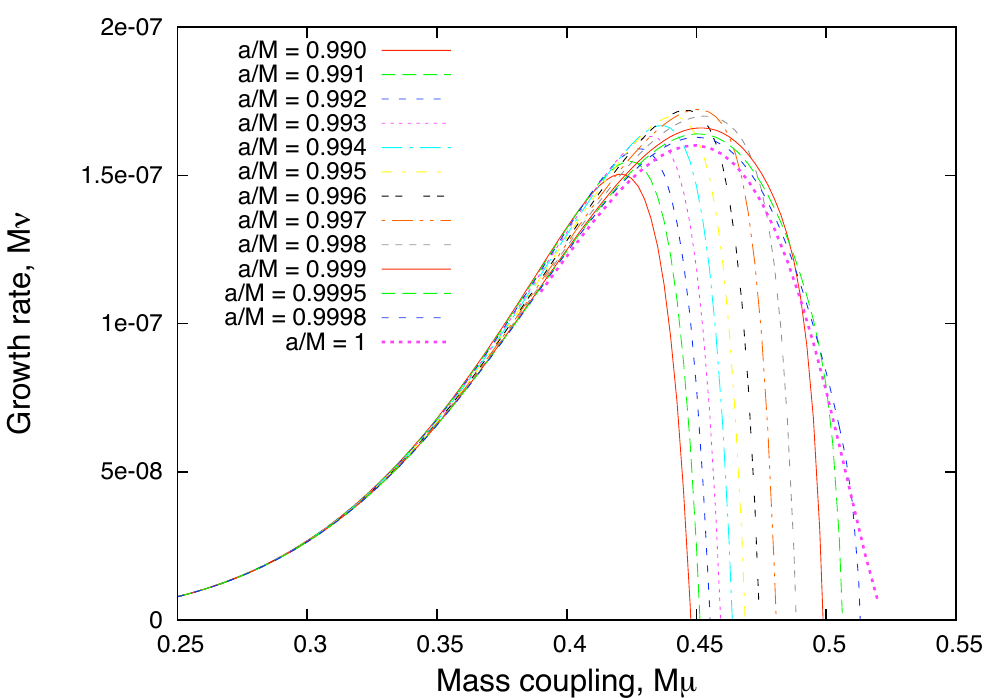}
 \caption{
 The growth rate $M \nu$ of the superradiant instability in the dominant mode of the scalar field ($l=m=1$, $n=0$), for rapidly-rotating black holes $0.99 \le a/M \le 1$, as a function of mass coupling $M\mu$. The growth rates for $a < M$ were found by computing the quasi-bound state spectrum with the continued-fraction method of Ref.~\cite{Dolan:2007}. The growth rates for the extremal case $a=M$ were found by direct integration of the radial equations. The plot shows that a maximal growth rate of $M \nu \approx 1.72 \times 10^{-7}$ occurs for $a \approx 0.997 M$ and $M \mu \approx 0.45$. 
  }
 \label{fig:freq-domain-growth}
\end{figure}

Figure \ref{fig:freq-domain-growth} shows the growth rate as a function of $M\mu$, for the dominant mode (the $l=m=1$, $n=0$ quasi-bound state) of rapidly-rotating black holes. The plot shows that the maximum possible growth rate is $M\nu_{\text{max}} \approx 1.72 \times 10^{-7}M$, corresponding to an e-folding time of $5.81 \times 10^{6} M$. Perhaps surprisingly, this maximum does \emph{not} arise at the extremal limit ($a = M$), but rather at around $a \approx 0.997M$ for $M \mu \approx 0.45$.
 
\begin{figure}
 \includegraphics[width=8.5cm]{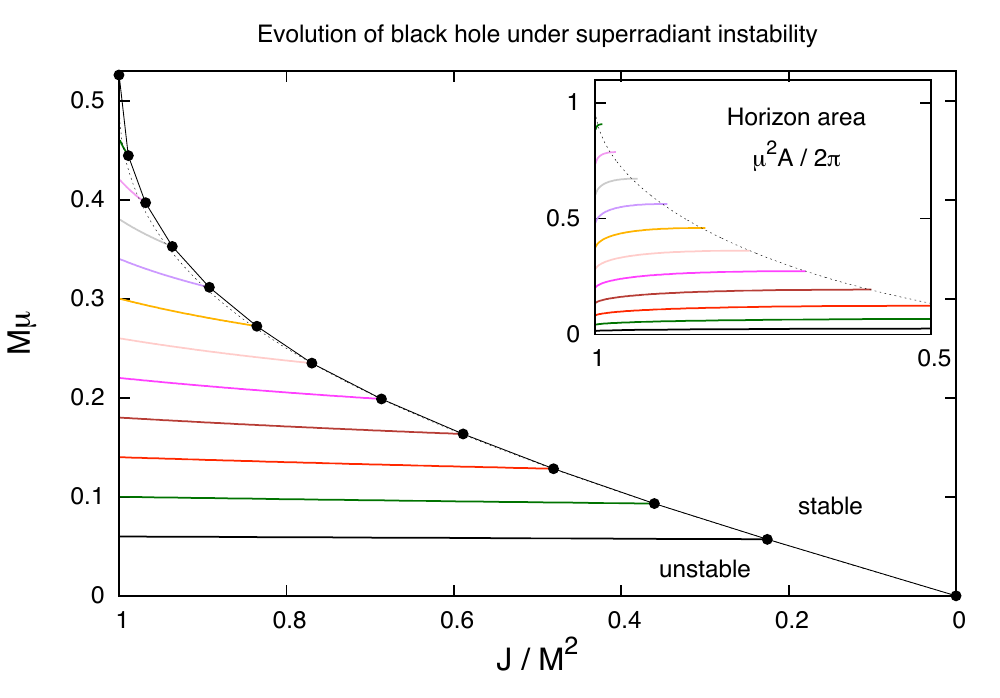}
 \caption{Evolution of black hole parameters under superradiant instability induced by a scalar field of mass $\mu$, in the linear approximation. In the superradiant regime, the black hole loses mass $M$ and angular momentum $J$ into the scalar field (predominantly emitting into the $m=l=1$, $n=0$ quasi-bound state). The BH parameters evolve along the coloured curves, from left to right (whereas the field mass $\mu$ is assumed to be constant). The instability ends at the marked points, which lie just above the (dotted) line $\mu = \Omega$. The inset, which shows $\mu^2 A / (2 \pi)$ as a function of $J / M^2$, illustrates that the horizon area $A$ \emph{increases} in this process, up to a maximum of $A \approx 4 \pi a \Omega / \mu^2$ (dotted line). 
 [N.B. Units $G=c=\hbar=1$ are used].
  }
 \label{fig:evolution}
\end{figure}

Under the instability, the black hole will shed mass and angular momentum into the dominant bound state, in the ratio $dJ / dM = m / \hat{\omega}$. Figure \ref{fig:evolution} illustrates the evolution of the black hole in the $M\mu$--$(J/M^2)$ plane. We make the simplifying assumptions that the process remains within the linear regime, and that the mass and angular momentum in the field are not re-absorbed by the BH as $E$ and $J$ shift. The plot illustrates that, once the instability is in progress, the BH will migrate from left to right along the coloured curves, spinning-down more rapidly than it loses mass, until it reaches the superradiant cut-off where $\hat{\omega} = \Omega$ ($\approx \mu$). The endpoint of the instability in the linear regime (i.e.~the modes with $\nu = 0$) was the focus of a recent work on the extremal ($a=M$) case \cite{Hod:clouds}. More realistic scenarios for the termination of the instability, taking into account non-linear effects, are the subject of recent studies \cite{Arvanitaki1, Arvanitaki2, Yoshino:Kodama}.
 

It should be noted that the concordant view of the instability, as developed in \cite{Furuhashi:2004, Cardoso:Yoshida:2005, Dolan:2007, Rosa:2010}, is challenged by two outlier studies. First, a study of the extremal Kerr BH \cite{Hod:Hod:2009a} predicted a maximum growth rate some four orders of magnitude faster than the consensus. However, it was shown in \cite{Rosa:2010} that the anomalous growth rate was likely due to a flaw in the matching method, because the functional forms used in the matching procedure were not valid across the full range of $M\mu$ (the functional forms exhibited spurious poles which led to incorrect estimates). Second, a study of the massive scalar field in the time-domain in 2005 \cite{Strafuss:2005} reported evidence for a fast growth rate, $M \nu \sim 2\times 10^{-5}$ (see Fig.~6 in Ref.~\cite{Strafuss:2005}). It was recently argued in Ref.~\cite{Witek} that this finding was in fact due to a misinterpretation of the `beating' effect, caused by interference between quasi-bound modes of similar frequencies, which leads to a slow modulation of the envelope of the field. 

\subsection{Mirror states\label{subsec:mirror}}
The relevant states in the case of a black hole surrounded by a mirror are defined by Eq.~(\ref{eq:ingoing}) and 
\beq
X(r_m) = 0
\eeq
where $r_m$ is the mirror radius. The mirrored case was considered in detail in Ref.~\cite{Cardoso:2004-bomb}. 

Once more, the states are labelled by angular momentum $l$, $m$ and overtone $n$. But, in contrast to the massive-field case, the number of overtones in the spectrum depends on the mirror radius $r_m$. The number of overtones $n_{\text{max}}$ increases with $r_m$, as more overtones `fit into' the cavity between horizon and mirror. 

The modes are approximately evenly-spaced in frequency, with a separation that decreases as the mirror radius increases. For moderate-to-large $r_m$, a simple approximation for the oscillation frequency $\hat{\omega}_n$ of the $n$th overtone is given by \cite{Cardoso:2004-bomb},
\beq
\hat{ \omega}_n \approx \frac{j_{l+1/2,n}}{r_m} \approx \frac{\left(n + 1 + l/2\right) \pi}{r_m} , \label{eq:mirrorspec}
\eeq 
where $j_{l+1/2,n}$ is the $(n+1)$th root of the Bessel function of the first kind $J_{l+1/2}(\cdot)$. Modes within the superradiant regime, $\hat{\omega}_n < m \Omega$, will grow with time, $\nu > 0$. An estimate of the growth rate is given in Eq.~(35)--(36) of Ref.~\cite{Cardoso:2004-bomb}. 
The growth rate of the mirrored system can be much larger than the massive-field case without mirror, as shown in Fig.~\ref{fig:mirror-rates}. This makes the mirrored case rather easier to observe in time-domain simulations.

\section{Time domain methods\label{sec:methods}}
Below we develop methods for solving the scalar-field wave equation in the time domain.

\subsection{2+1D equation}
We begin by separating out the azimuthal dependence of the field, 
\beq
 \Phi = \frac{1}{r} \Psi (t,r,\theta) e^{im\varphi}
\eeq
after introducing an alternative azimuthal coordinate $\varphi$ via
\beq
d\varphi = d \phi + \frac{a}{\Delta} dr .
\eeq
The new coordinate is introduced to handle the problem that the Boyer-Lindquist coordinate $\phi$ is undefined as the horizon is approached (equivalently, to remove a term in the potential, $a^2 m^2$, which does not vanish as $r \rightarrow r_+$). This ansatz leads to a 2+1D equation,
\begin{eqnarray}
\left\{ - \Sigma^2 \partial_{tt} + (r^2+a^2)^2 \partial_{xx} - 4iamMr \partial_{t} \right. && \nn \\
\left. + \left[ 2 i a m (r^2+a^2) - 2a^2\Delta / r \right] \partial_x - \hat{V} \right\} \Psi &=& 0 ,  \label{eq:2+1}
\end{eqnarray}
where $\Sigma^2 = [(r^2 + a^2)^2 - a^2 \Delta]  + a^2 \Delta \cos^2 \theta$, and $\partial_x$, $\partial_{xx}$ denote 1st and 2nd partial derivatives with respect to the tortoise coordinate, and
\begin{eqnarray}
\hat{V} &=& \Delta \left[ -\left\{ \partial_{\theta \theta} +\cot \theta \partial_\theta - \frac{m^2}{\sin^2\theta} \right\} + \frac{2M}{r}\left(1 - \frac{a^2}{Mr} \right) \right. \nn \\
 && \left. \quad + \frac{2iam}{r} + (r^2 + a^2 \cos^2 \theta) \mu^2 \right] .
\end{eqnarray}
Eq.~(\ref{eq:2+1}) can be solved with standard finite-difference methods, as in Refs.~\cite{Strafuss:2005, DWB}. Indeed, in Sec.~\ref{subsec:validation} we make use of a well-tested 2+1D code to check our implementation. 

\subsection{Coupled 1+1D equations}
Let us now make the further decomposition,
\beq
\Psi = \sum_{j=|m|}^{\infty} \psi_{j}(r) Y_{jm}(\theta) ,  \label{eq:lmode}
\eeq
where $Y_{jm}(\theta)$ denotes a standard spherical harmonic without the azimuthal part. Now we substitute (\ref{eq:lmode}) into Eq.~(\ref{eq:2+1}) and, by projecting onto the basis of harmonics, we recover a set of coupled 1+1D equations. The $\cos^2\theta$ terms in Eq.~(\ref{eq:2+1}) lead to some coupling between $l$ modes; however, the coupling turns out to be rather straightforward. We make use of the following result \cite{Hughes:2000},
\begin{eqnarray}
\left< l m | \cos^2\theta | j m \right> &=& \frac{1}{3} \delta_{jl} + \frac{2}{3} \sqrt{\frac{2 j + 1}{2 l + 1}} \left< j, 2, m, 0 | l, m \right> \times \nn \\
 && \quad  \left<j, 2, 0, 0 | l, 0\right> \equiv c_{jl}^m
\end{eqnarray}
where 
\beq
\left< l m | \mathcal{X}(\theta) | j m\right> = 2 \pi \int  Y^\ast_{l m}(\theta) \mathcal{X}(\theta) Y_{jm}(\theta) d(\cos \theta) .
\eeq
The numbers $\left< j,i,m,0|l,n \right>$ are Clebsch-Gordan coefficients. The coupling coefficient $c_{jl}^m$ is zero unless $j = l-2$, $l$ or $l+2$. Hence, the even-$l$ and odd-$l$ sectors are completely decoupled (as expected from symmetry considerations), and, within each sector, the coupling between $l$-modes of the same parity is nearest-neighbour only.

Let us now write the equations in first-order-in-time form by introducing $\pi_l = \partial_t \psi_l$, so that
\begin{eqnarray}
&&\left[ \Sigma^2_{(0)} + a^2 \Delta c_{ll}^m \right] \dot{\pi}_{l}  + a^2 \Delta \left( c_{l,l+2}^m \dot{\pi}_{l+2} + c_{l,l-2}^m \dot{\pi}_{l-2} \right) = \nn \\
&&(r^2+a^2)^2 \psi_l^{\prime\prime} + \left(2iam(r^2+a^2) - 2a^2\Delta/r\right) \psi_l^\prime \nn \\ 
&& - 4iamMr \pi_{l} - V_l - a^2 \mu^2 \Delta \left( c_{l,l+2} \psi_{l+2} + c_{l,l-2} \psi_{l-2} \right) \nn \\
  \label{eq:1+1-coupled}
\end{eqnarray}
where $\Sigma^2_{(0)} = (r^2+a^2)^2 - a^2 \Delta$ and 
\begin{eqnarray}
V_l &=& \Delta \left[ l(l+1) +\frac{2M}{r} \left(1-\frac{a^2}{Mr}\right) + \frac{2iam}{r}  + \right. \nn \\
  && \quad \quad \left. \mu^2 (r^2 + a^2 c_{ll}^m) \right] \psi_l .
\end{eqnarray}

\subsection{Stress-energy\label{sec:stress-energy}}
The field equation (\ref{eq:kg}) is obtained from an action principle with a Lagrangian $\Lag$, where
\beq
\Lag = -\tfrac{1}{2} g^{\alp \bet} \partial_{(\alp} \Phi^\ast \partial_{\bet)} \Phi - \tfrac{1}{2} \mu^2 \Phi^\ast \Phi .
\eeq
The associated stress-energy tensor $\tensor{T}{^\alp _\bet}$ is
\beq
\tensor{T}{^\alp _\bet} = -g^{\alp \gam} \partial_{(\bet} \Phi^\ast \partial_{\gam)} \Phi - \delta^\alp_\bet \Lag ,
\eeq
where $X_{(\alp}Y_{\bet)} \equiv \tfrac{1}{2}\left(X_\alp Y_\bet + Y_\alp X_\bet \right)$. The stress-energy is symmetric in its indices, $T_{\alp \bet} = T_{\bet \alp}$, and locally conserved, $\tensor{T}{^\alp ^\bet _{;\bet}} = 0$. By contracting the stress-energy with a Killing vector of the spacetime we may form a conserved vector field. There are two linearly-independent Killing vectors on Kerr spacetime, and we may form vectors associated with energy and azimuthal angular momentum in this way. By constructing a four-volume and applying Gauss' theorem (as in e.g.~Sec.~IIB of Ref.~\cite{Dolan:2007}), we obtain two conservation laws in the following form,
\beq
\frac{dE}{dt} = - \FE, \quad \frac{d\J}{dt} = - \FJ.  \label{eq:cons-law}
\eeq
The `energy' $E$ and `azimuthal angular momentum' $\J$ are found from integrals over a timelike hypersurface. If we choose an `exterior' region $x \ge \xinner$ of a constant-$t$ 3-surface, we may write
\beq
E = \int_{\xinner}^\infty \mathcal{E} dx , \quad \J = \int_{\xinner}^\infty \mathcal{J}_z dx.   \label{eq:EJ}
\eeq
Here $\mathcal{E}$ and $\mathcal{J}_z$ are the energy and angular momentum densities, given by
\beq
\mathcal{E} = \sum_{l} \mathcal{E}_l, \quad \mathcal{J}_z = \sum_{l} \mathcal{J}_l,
\eeq
where
\begin{eqnarray}
&& 2 (r^2+a^2) \mathcal{E}_{l}=  \Delta \left[ l(l+1) + \mu^2 (r^2 + a^2 c_{ll}^m) \right] \left|\Phi_l\right|^2 \nn \\
&& \quad + 2 \Delta a^2 \mu^2 c_{l,l+2}^m \text{Re}\left( {\Phi_l^\ast \Phi_{l+2}} \right) \nn \\
&& \quad + (\Sigma^2_{(0)} + \Delta a^2 c_{ll}^m) \left| \dot{\Phi}_l \right|^2 + 2 \Delta a^2 c_{l,l+2}^m \text{Re} \left(\dot{\Phi}_l^\ast \dot{\Phi}_{l+2} \right) \nn \\
&& \quad + (r^2+a^2)^2 \left| \Phi_l^\prime \right|^2 + 2 a m (r^2+a^2) \text{Im}\left( \Phi_l^\ast \Phi_l^\prime \right) , \label{eq:Edensity} 
\end{eqnarray}
and 
\begin{eqnarray}
&&-(r^2+a^2) \mathcal{J}_l = \left( \Sigma_{(0)}^2 + a^2 \Delta c_{ll}^m \right) m \text{Im}\left( \dot{\Phi}_l \Phi^\ast_l \right) \nn  \\
&&\quad + m a^2 \Delta c_{l,l+2}^m \text{Im} \left( \dot{\Phi}_{l+2} \Phi^\ast_l + \dot{\Phi}_l \Phi^\ast_{l+2} \right) \nn \\
&&\quad + 2 M a r m^2 \left| \Phi_{l} \right|^2,  \label{eq:Jdensity}
\end{eqnarray}
where $\Phi_l = r \psi_l$, $\dot{\Phi}_l = \partial_t \Phi_l$, $\Phi^\prime_l = \partial_x \Phi_l$, and $\tilde{\Phi}^\prime_l = \Phi^\prime_l + i a m \Phi_l / (r^2+a^2)$.

The fluxes are found from taking surface integrals over the spacelike surface at $x = \xinner$, leading to
\begin{eqnarray}
\FE &=& (r^2+a^2) \sum_{l} \text{Re} \left( \dot{\Phi}_{l}^\ast \tilde{\Phi}_{l}^\prime \right) ,  \label{eq:fluxE} \\
\FJ &=& -m (r^2 + a^2) \sum_{l} \text{Im} \left( \Phi_{l}^\ast \tilde{\Phi}_{l}^\prime \right) .  \label{eq:fluxJ}
\end{eqnarray}

Let us now consider a wave of frequency $\omega$ which is `ingoing' in the near-horizon region, i.e.~$\Phi_l \sim A_l \exp(-i\omega (t+x))$. Such a wave generates the fluxes $\FE \sim 2 M r_+ \left|A_l\right|^2 \omega \tilde{\omega}$ and $\FJ \sim (m / \omega) \FE$. These fluxes become negative in the superradiant regime $0 < \omega < m\Omega$. Hence, though the wave is moving inwards, energy and angular momentum are passing outwards into the exterior region.

The choice of inner boundary $\xinner$ is somewhat arbitrary (typically, we take $\xinner = -100M$), and necessitated by the use of a finite grid. Nevertheless, the idea of `energy in the exterior spacetime' is a useful physical concept; we will see that, by tracking the energy and AM in the exterior, we may track the development of the instability. We note that shifting the boundary $\xinner$ merely has the effect of shifting the time-origin in the plots of $E$ and $J_z$, providing that $\xinner$ remains in the near-horizon region ($\xinner / M \ll 0$).

The conservation laws (\ref{eq:cons-law}) also provide a useful code test, because the following quantities (representing total energy and angular momentum) should remain constant in our evolution scheme, up to numerical error:
\beq
\CE = E + \int_{0}^t \FE dt, \quad \CJ = J_{z} + \int_{0}^t \FJ dt .  \label{SE-constants}
\eeq

\subsection{Finite-difference method\label{subsec:FD}}

We evolve the equations on a uniform grid with grid spacing $\Delta x = M / n$ for $n = 4$ to $16$. We use  time steps $\Delta t = \kappa \Delta x$ that are sufficiently small to satisfy the Courant-Friedrichs-Lewy stability condition, i.e.~with $\kappa = 0.8$,  typically. We use the Method of Lines, with a fourth-order Runge-Kutta time step, and fourth-order finite differencing on spatial slices, i.e.
\begin{eqnarray}
\psi_{(i)}^{\prime} &\approx& [ -\psi_{(i+2)} + 8 \psi_{(i+1)} - 8 \psi_{(i-1)} + \psi_{(i-2)} ] / (12 \Delta x) , \nn \\
\psi_{(i)}^{\prime \prime} &\approx& [ -\psi_{(i+2)} + 16 \psi_{(i+1)}  - 30 \psi_{(i)} \nn \\ 
&& \quad \quad \quad \quad + 16 \psi_{(i-1)} - \psi_{(i-2)} ] / (12 \Delta x^2)
\end{eqnarray}
where $\psi_{(i)}$ denotes $\psi_{j}(x = \xinner + i\Delta x)$ (and here $i$ is an integer).

The equation set (\ref{eq:1+1-coupled}) resembles
\beq
\mathbf{A} \dot{\mathbf{x}} = \mathbf{b}   \label{eq:matrix}
\eeq
where $\mathbf{x}^T = \left[ \pi_{k}, \pi_{k+2}, \pi_{k+4}, \ldots \right]$ (where $k = |m|$ or $|m|+1$, depending on parity) and $\mathbf{A}$ is a real, symmetric, diagonally-dominant, tridiagonal matrix.  At each grid point, we solve (\ref{eq:matrix}) to find $\dot{\mathbf{x}}$. Fortunately, as $\mathbf{A}$ is tridiagonal, the solution can be found in a numerically-efficient way, using (e.g.)~the Thomas algorithm \cite{NumericalRecipes}.

\subsection{Perfectly matched layers}
To achieve computational efficiency, it is necessary to truncate the computational domain. As all physical solutions are `ingoing' in the near-horizon region, one possibility is to impose the ingoing condition $\left( \partial_t - \partial_x  \right) \psi_l = 0$ on finite-difference molecules at the left boundary of the grid. However, this approach -- imposing a condition at an interface -- runs the risk of generating high-frequency artifacts, which can excite numerical instabilities, and which may be mistaken for the superradiant instabilities themselves. In this work, we take an alternative approach: we truncate the computational domain by introducing an artificial absorbing region, which smoothly attentuates the solution without generating reflections. This is the concept of the `perfectly matched layer' (PML) \cite{Berenger:1994}.

PMLs are widely used in computational electrodynamics \cite{Taflove:Hagness}, but the underlying idea can be applied to a range of wave equations. For example, PMLs have previously been used in simulations of `analogue' black holes in Bose-Einstein condensates \cite{Farrell:Leonhardt, Giovanazzi}. 

We may illustrate the PML approach with the simple example of the 1D wave equation, $\ddot{u} = u^{\prime\prime}$. This may be written in fully first-order form as $\dot{u} = v^\prime$, $\dot{v} = u^\prime$. In the frequency domain, it has wave solutions $\exp(-i \omega (t \pm x))$. To create a PML, we may analytically-continue the wave solutions onto a complex-$x$ contour and then make a coordinate transformation to express $x$ as a function of a real coordinate. These steps are equivalent to making the replacements
\beq
\frac{\partial}{\partial x} \rightarrow \frac{1}{1 + i \pml(x) / \omega} \frac{\partial}{\partial x} ,
\eeq
in the frequency-domain equations. In the time-domain equations this leads to
\begin{eqnarray}
\dot{u} &=& v^\prime - \pml(x) u , \label{u-pml} \\
\dot{v} &=& u^\prime - \pml(x) v , \label{v-pml}
\end{eqnarray}
the solutions of which are given by
\beq
u \sim \exp\left(- i \omega (t \pm x) \pm \int^x_{y} \pml(x) dx  \right) .
\eeq
and $v = \pm u$. It follows that a right-going wavepacket moving into a PML from the left (i.e.~with $y < x$) will be attentuated; and also that a left-going wave moving into a PML from the right (i.e.~with $y > x$) will also be attentuated. Note that here $\pml$, the amplitude of the attentuation, is a function of $x$ that may be varied smoothly, over the scale of a few wavelengths, without generating reflections (at the continuum level). Thus we may `turn on' the PML in a smooth manner, which greatly reduces the risk of numerically-generated high-frequency reflections. 

How may this idea applied to the wave equation on Kerr? First, we note that in the near-horizon region $x / M \ll 0$, the 1+1D equations decouple, reducing to
\beq
\ddot{\psi}_l - \psi_l^{\prime\prime} + 2im \Omega \dot{\psi}_l - 2im \Omega \psi_l^\prime = 0 .
\eeq
This equation may be turned into a simple 1D wave equation, $\ddot{u} - u^{\prime\prime} = 0$ by making the replacement $\psi_l = \exp(- i m \Omega (t + x) ) u$. The PML modification can then be applied, at the expense of introducing an auxiliary variable to play the role of $v$. 
After transforming back to the original variables we are led to
\begin{eqnarray}
\dot{\psi} &=& \pi , \nn \\
\dot{\pi} &=& \psi^{\prime \prime} + 2 i m \Omega (\psi^\prime - \pi) - \chi^\prime - i m \Omega \chi - \pml \left( \pi + i m \Omega \psi \right) , \nn \\
\dot{\chi} &=& \pml \left( \psi^\prime - \chi + i m \Omega \psi \right) - i m \Omega \chi ,
\end{eqnarray}
(where the subscript $l$ has been dropped for clarity). To get the $\dot{\pi}$ equation, we combine the time-derivative of (\ref{u-pml}) with the spatial-derivative of (\ref{v-pml}) and introduce the modified auxiliary variable via $\chi = \gamma \exp(- i m \Omega (t + x) ) v$. The $\dot{\chi}$ equation follows from (\ref{v-pml}).

 Note that these equations only apply in the near-horizon region where terms $\sim \Delta$ are small enough to be neglected; this is the region where we add the Perfectly Matched Layer. Note also that $\chi$, the auxiliary variable, does not propagate outside the PML region.

\subsection{Grid boundaries}
Even with the use of PML, it is still necessary to take steps to handle the boundaries of the grid. We make use of one-sided spatial derivatives in the finite-difference molecules near the left and right boundaries. On the left boundary, beyond the PML region, we imposed a simple zero condition $\psi = 0 = \pi$ on the final grid point. Although this simple condition creates reflections, it merely reflects attenuated flux back into the PML region, where it is attenuated once more, to a numerically-negligible level. On the right boundary, we also impose a simple zero condition on the final grid point. In the mirror case, this is physically well-motivated (see Sec.~\ref{subsec:mirror}). In the massive-field case, we place the right boundary far from the BH, typically at $r_{\text{max}} \sim 500M$. As long as the boundary is far from the black hole, we do not expect it to significantly influence the lowest quasi-bound states, or the superradiant instability, as Ref.~\cite{Cardoso:2004-bomb} suggests that $M \nu \sim (M/r_{\text{max}})^4$ (for dipole of the massless field), so $M \nu \sim 1.6 \times 10^{-11}$ for $r_\text{max} = 500M$. We have carefully checked that the observed instability growth rate is not sensitive to the position of the right boundary. 


\subsection{Validation\label{subsec:validation}}
\begin{figure}
 \includegraphics[width=8cm]{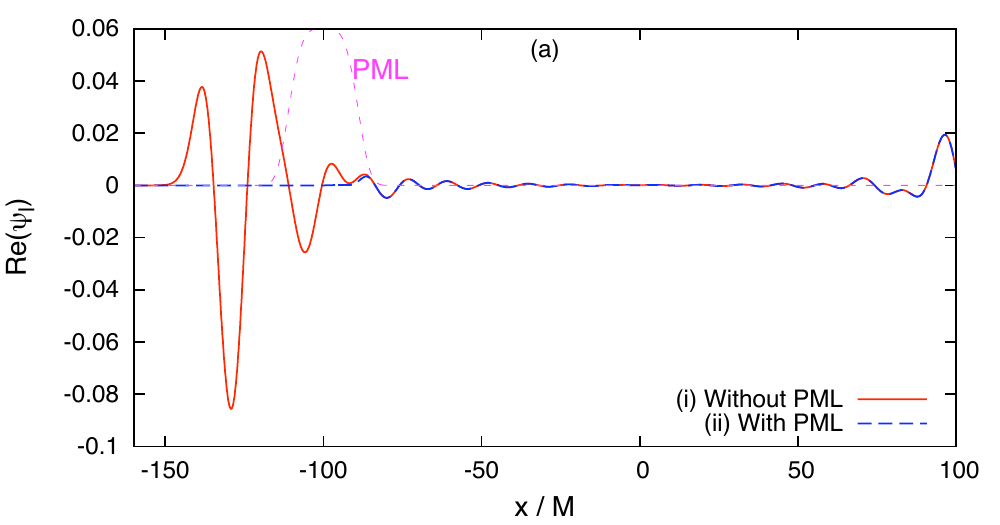}
 \includegraphics[width=8cm]{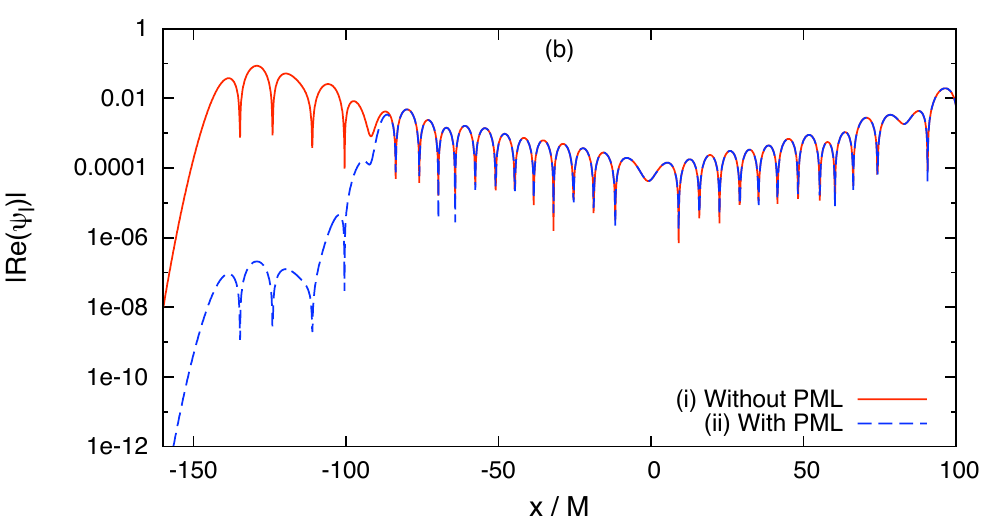}
 \vspace{0.1cm}
 \includegraphics[width=8cm]{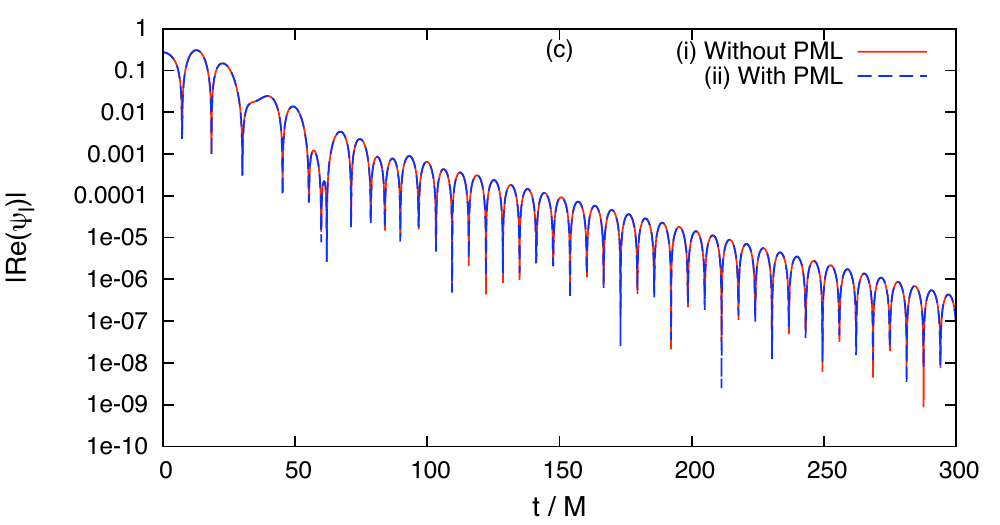}
 \caption{Validation of the Perfectly-Matched Layer method. Plot (a) compares snapshots taken at $t=140M$, from two evolutions of the same initial data (a time-symmetric Gaussian of width $\xwid = 5M$ centred around $\xmid = 10M$ in mode $l=m=1$) for a BH of spin $a=0.99M$, using (i) [red, solid] no PML, and (ii) [blue, dashed] a PML centred around $x = -100M$ of amplitude $\pml(x)$ (with $\pml / 10$ shown as dashed magenta line).  Outside the PML region, (i) and (ii) are in good agreement. Inside the PML region, incident waves are smoothly attenuated without reflection. Plot (b) shows the same data on a logarithmic scale.  Plot (c), showing data extracted at $x=x_1=0$ as a function of time, illustrates that, to the right of the PML, data sets (i) and (ii) are virtually indistinguishable.  }
 \label{fig:pml}
\end{figure}

We have made a number of checks of the correctness of our implementation. First, we tested the numerical convergence of the finite-different code via a ratio test, and found strong evidence for the expected rate of convergence. Second, we compared the results of the 1+1D code with those of the well-validated 2+1D code, finding agreement up to the anticipated level of numerical error. Third, we tested the implementation of the PML, by comparing two data sets arising from evolving the same initial data (i) on a domain in $x$ without a PML and sufficiently large that boundaries are not encountered, and (ii) on a restricted domain, $x \ge -200M$, with a PML of width $10M$ centred around $x = -100M$. Figure \ref{fig:pml} shows typical results; data sets (i) and (ii) are found to be in excellent agreement on the right-hand side of the PML, as expected. Fourth, we have checked that the `quasinormal ringing' has the expected frequency and decay time. Fifth, we have checked that stress-energy `constants' $C_{(E/J)}$ [Eq.~(\ref{SE-constants})] are preserved in our runs, up to numerical error (see Fig.~\ref{fig:psi-mu}). Sixth, we have checked that the results presented here are robust under adjustments of `internal' parameters (such as the width and amplitude of the PML; the maximum number of $l$ modes used; and the grid resolution). Finally, the next sections show excellent agreement with frequency-domain results, which we take to be further evidence in favour of a valid implementation.

\section{Results\label{sec:results}}
In this section we present a selection of numerical results, for two cases in which an instability is present: the massless scalar field with a mirror (Sec.~\ref{subsec:mirror}), and the massive scalar field without mirror (Sec.~\ref{subsec:massive}). In each case, we seek to stimulate the system with some broadband initial data -- typically a Gaussian in $x$ -- and study the features of the response. In particular, we attempt to compute the spectrum of superradiant quasi-bound states which dominate the response at late times. We will focus on the $m=1$ mode, on which superradiance is expected to have the strongest effect.

\subsection{Scalar field with mirror\label{subsec:mirror}}

\begin{figure*}
 \includegraphics[width=14cm]{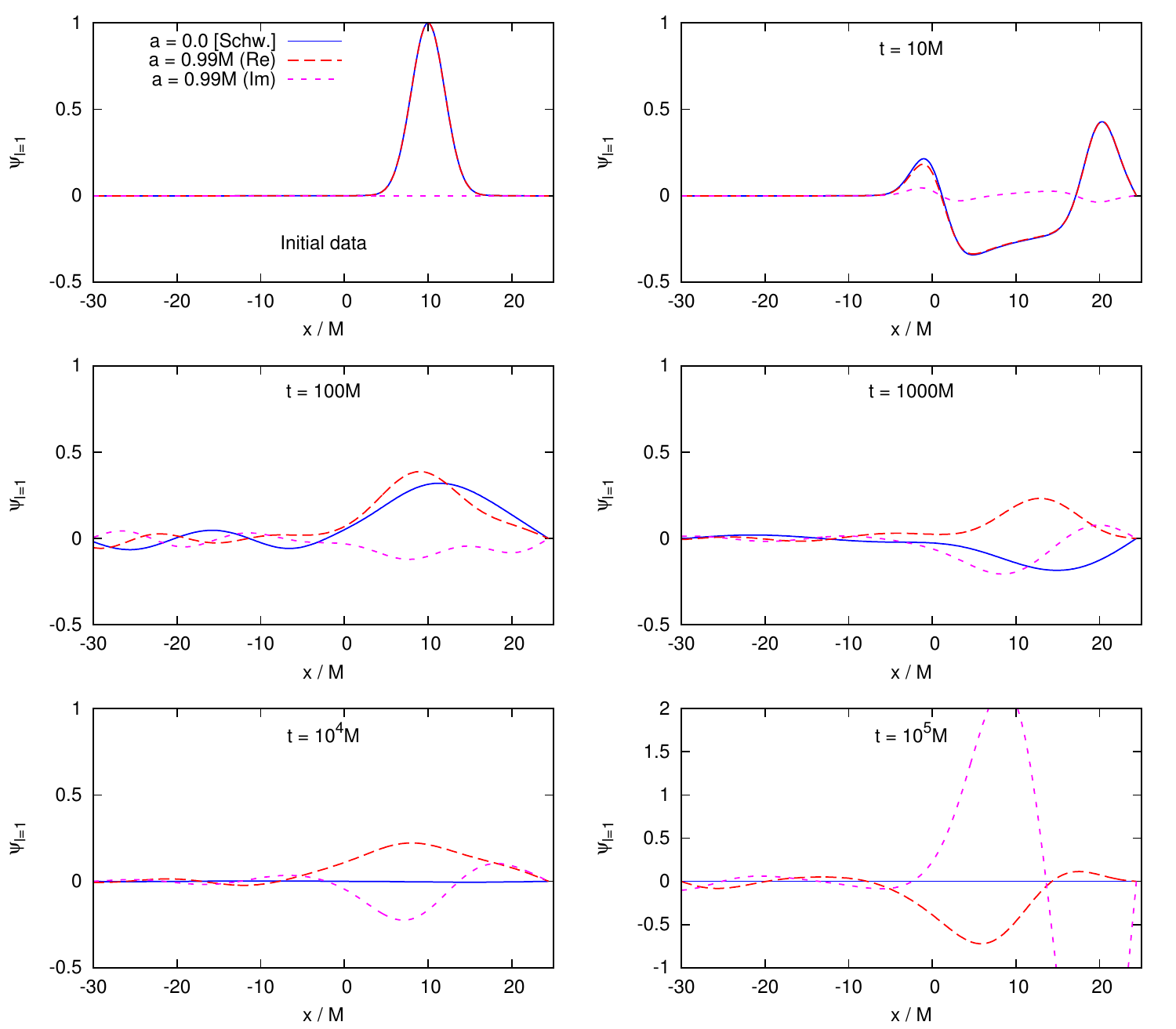}
 \caption{Time evolution of field on black hole background with `mirror' at $r_m = 20M$. These plots show snapshots of $\psi_l$ [Eq.~(\ref{eq:lmode})] for the dipole mode $l = m = 1$ of the scalar field, as a function of tortoise coordinate $x/M$, at $t = 0$ (top left), $10M$, $100M$, $10^3M$, $10^4M$ and $10^5M$ (bottom right), for the Schwarzschild ($a=0$, solid line) and fast-rotating Kerr ($a = 0.99M$, broken lines) BHs.  Here the initial data is a time-symmetric Gaussian (\ref{eq:gaussian}) of width $\xwid = 2M$ centred about $\xmid = 10M $. By late times, the field has been absorbed by the Schwarzschild BH, and amplified by the Kerr BH.}
 \label{fig:psi-snaps}
\end{figure*}

Let us begin by considering a BH surrounded by a mirror at $r_m = 20M$, subject to some smooth initial perturbation. Figure \ref{fig:psi-snaps} compares the radial profile of the response of a Schwarzschild BH ($a = 0$) with that of a rapidly-rotating Kerr BH ($a = 0.99M$), over several decades of simulation time $t/M = 10, 100, \ldots 10^5$. Here the initial data is a time-symmetric Gaussian, 
\beq
\psi_{l=1} = \exp\left(-(x-\xmid)/[2\xwid^2] \right), \quad \psi_{l>1} = 0 = \dot{\psi}_{l} ,  \label{eq:gaussian}
\eeq
with $\xmid = 10M$ and $\xwid = 2M$. At early times (e.g.~$t=10M$), the response of the two BHs is very similar. At intermediate times, $t=100M$--$1000M$, some part of the initial perturbation has been absorbed through the horizon, and some remains in the exterior; the responses of Schwarzschild and Kerr are now quite different, but the field amplitudes remain comparable. At late times, $t = 10^4M$ and $t = 10^5M$ we see that the scalar field on the Schwarzschild BH has essentially disappeared through the horizon. On the other hand, the scalar field on Kerr has experienced significant amplification (note vertical scale). 

\begin{figure*}
 \includegraphics[width=14cm]{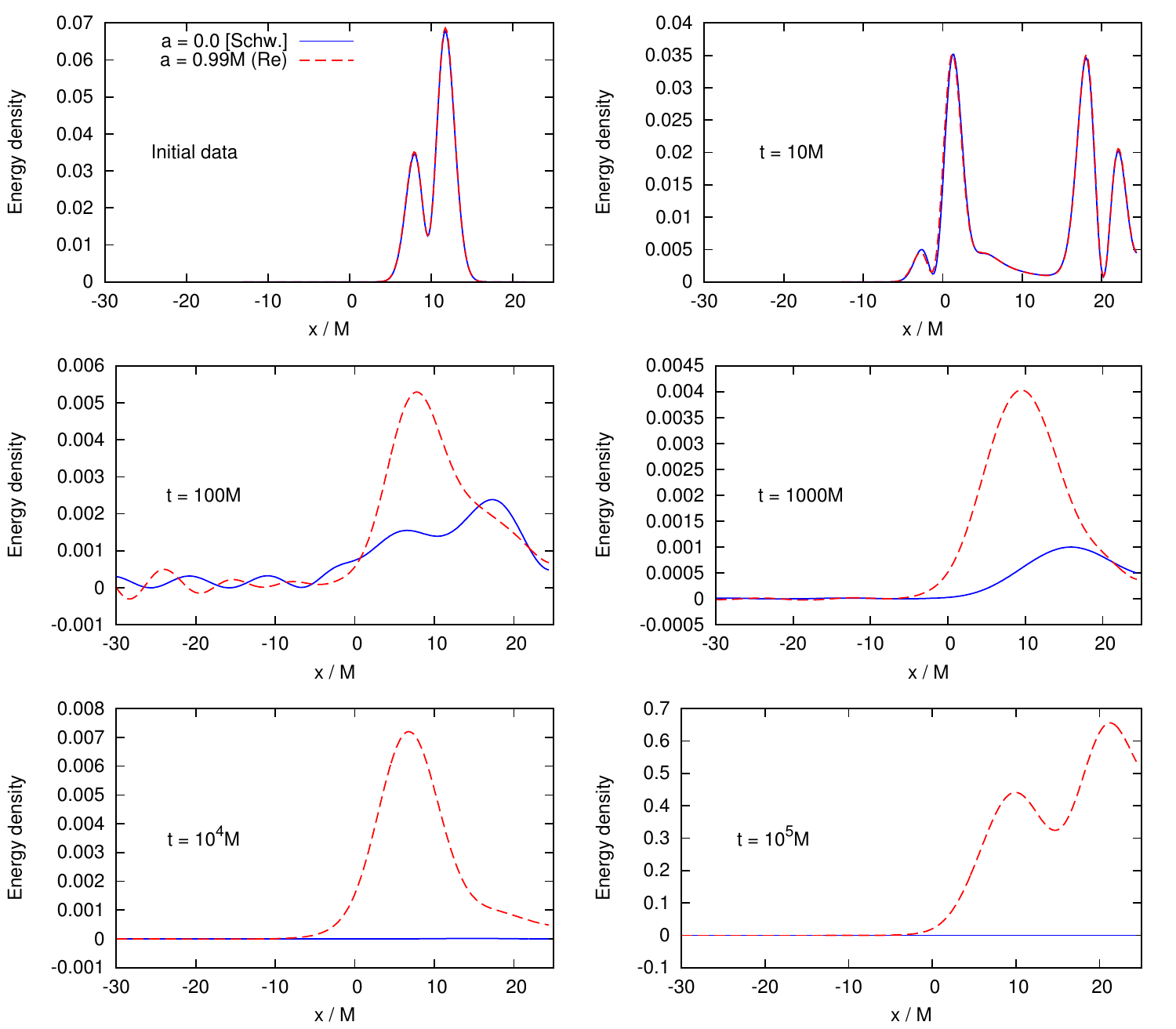}
 \caption{Time evolution of energy density $\mathcal{E}$ on black hole spacetime with `mirror' at $r_m = 20M$. These plots show snapshots of the energy density obtained from the field shown in Fig.~\ref{fig:psi-snaps} (see caption) via Eq.~(\ref{eq:Edensity}). The last plot (note the scale) shows that the energy of the field on the rotating BH (broken line) has increased by a factor of $\sim 100$ over a timescale of $\sim 10^5M$ via the superradiant mechanism; whereas the field on the non-rotating spacetime ($a=0$) has drained away through the horizon.}
 \label{fig:T00-snaps}
\end{figure*}

These conclusions are supported by Fig.~\ref{fig:T00-snaps}, which shows the radial profile of the energy density $\mathcal{E}$ [Eq.~(\ref{eq:Edensity})] computed from the field shown in Fig.~\ref{fig:psi-snaps}. The late time plots suggest that there is more than one growing mode in the Kerr case, and that the overtone with a peak closest to the mirror surface is emerging to dominate the signal.

\begin{figure}
 \includegraphics[width=8cm]{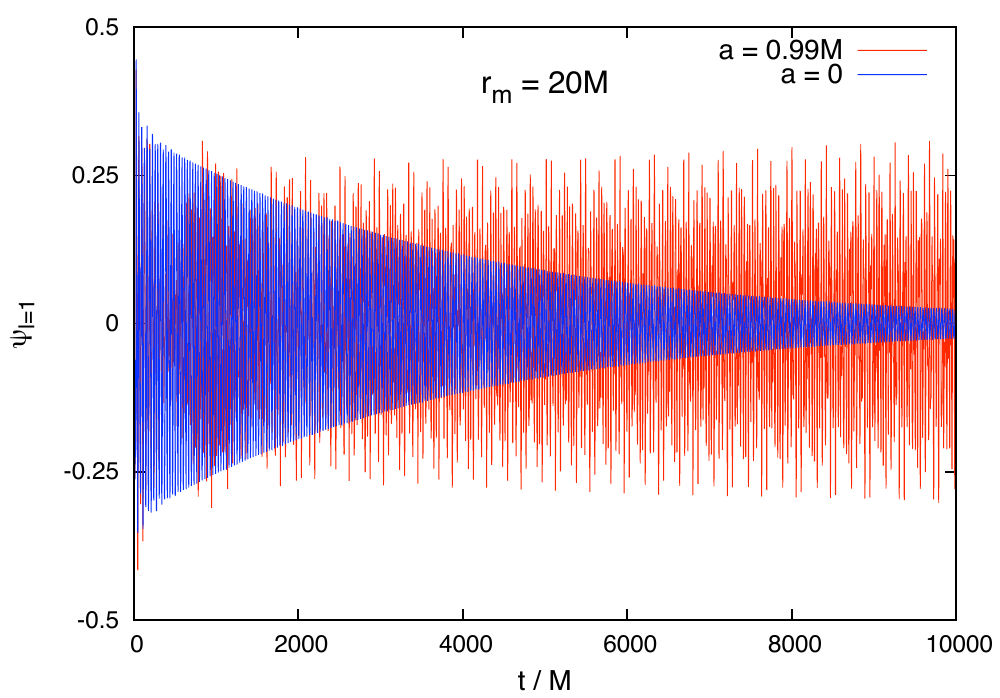}
 \caption{Time series data for the evolution of the field on a BH with mirror. The plot shows the (real part of) the field shown in Fig.~\ref{fig:psi-snaps} and \ref{fig:T00-snaps} extracted at $r = 10M$, as a function of time $t/M$. On the non-rotating spacetime ($a=0$, blue) the field decays, whereas on the rotating spacetime ($a=0.99M$, red) it is long-lived. The latter case shows evidence of `beating', due to interference between modes of different frequencies, and growth (see Fig.~\ref{fig:timeseries2}).}
 \label{fig:timeseries1}
\end{figure}

\begin{figure}
 \includegraphics[width=8cm]{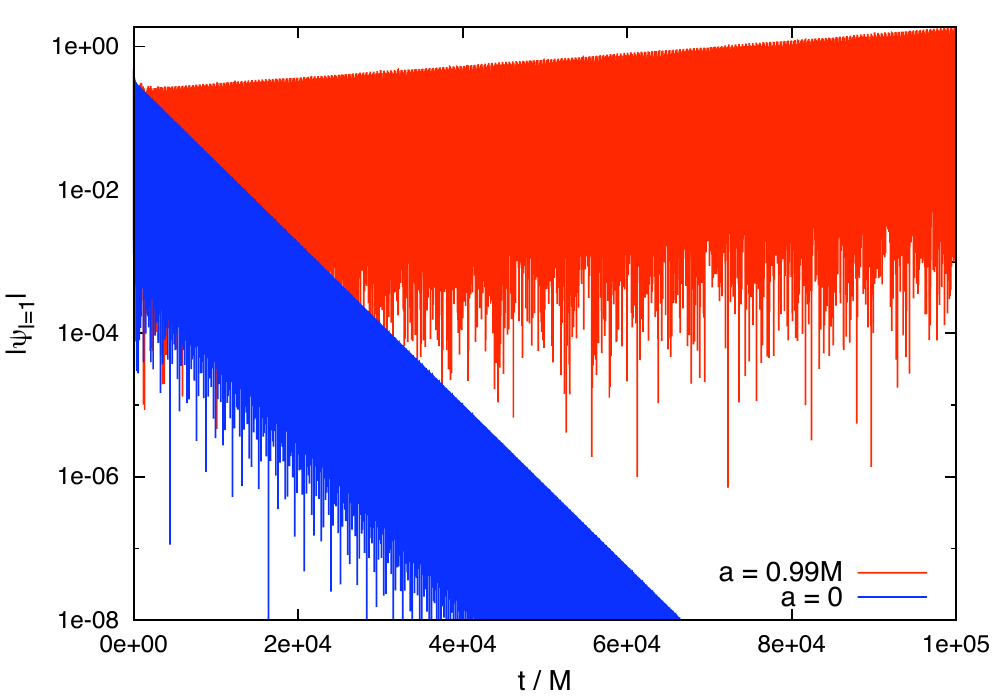}
 \includegraphics[width=8cm]{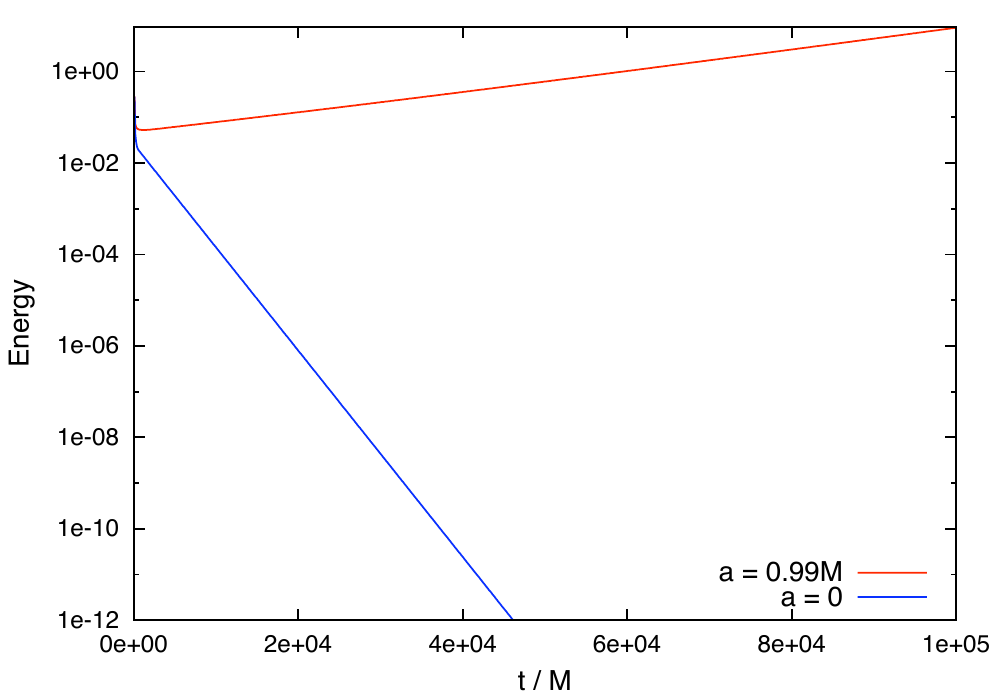}
 \caption{These plots show the time evolution of the field as in Fig.~\ref{fig:timeseries1}, on a logarithmic scale, up to $t = 10^5M$. The upper plot shows that the envelope of the Kerr (Schwarzschild) field grows (decays) exponentially. The lower plot shows that the energy in the exterior (see Sec.~\ref{sec:stress-energy}) also grows (decays) exponentially without beating effects, and with twice the exponent, as expected. }
 \label{fig:timeseries2}
\end{figure}

Let us now consider the field measured at a fixed point $x = x_1$ as a function of time. Figures \ref{fig:timeseries1} and \ref{fig:timeseries2} (log scale) show typical time series data for Schwarzschild and Kerr BHs. In the Schwarzchild case, the field decays exponentially, as the mirror reflects flux back onto the absorbing horizon. In general, the decay rate increases as the mirror is moved closer to the BH. In the Kerr case, the field is long-lived. Figure \ref{fig:timeseries1} shows that there is typically a `beating' effect in the signal, generated by interference between modes of different frequencies. This leads to a modulation of the signal which can disguise the underlying growth and even lead to incorrect conclusions (as argued in Ref.~\cite{Witek}). On the other hand, Fig.~\ref{fig:timeseries2} shows that the total energy in the exterior $E$ (Sec.~\ref{sec:stress-energy}) gives a clear view of the underlying growth as: (i) $E$ does not oscillate sinusoidally, and (ii) the variation of $E$ depends only on the energy flux through the inner boundary, $\FE$ [Eq.~(\ref{eq:fluxE})], which is only marginally influenced by beating between modes. 

\subsubsection{Mirror spectrum\label{subsec:mirrorspec}}
The spectrum of the response is revealed by taking a Fast Fourier Transform of the time series data for the field (extracted from the grid at fixed $x = x_1$). Let us introduce the (complex) Fourier amplitude, defined by
\beq
f_l(\omega_j) = \frac{1}{N} \sum_{k=0}^{N-1} \psi_{l}(t_k) \exp(-i \omega_j t_k ) ,
\eeq
where $t_{k} = k \Delta t$, $\omega_j = j \Delta \omega$, and $N$ is the number of data points, $\Delta t = t_{\text{max}} / (N-1)$ and $\Delta \omega = 2 \pi / t_{\text{max}}$, and the power spectrum given by
\beq
P_l(\omega) = \left| f_l(\omega) \right|^2 .  \label{eq:power}
\eeq
Note that the frequency resolution $\Delta \omega$ is inversely proportional to the simulation time $t_{\text{max}}$; hence longer runs yield a more refined view of the spectrum. 

\begin{figure*}
 \includegraphics[width=10cm]{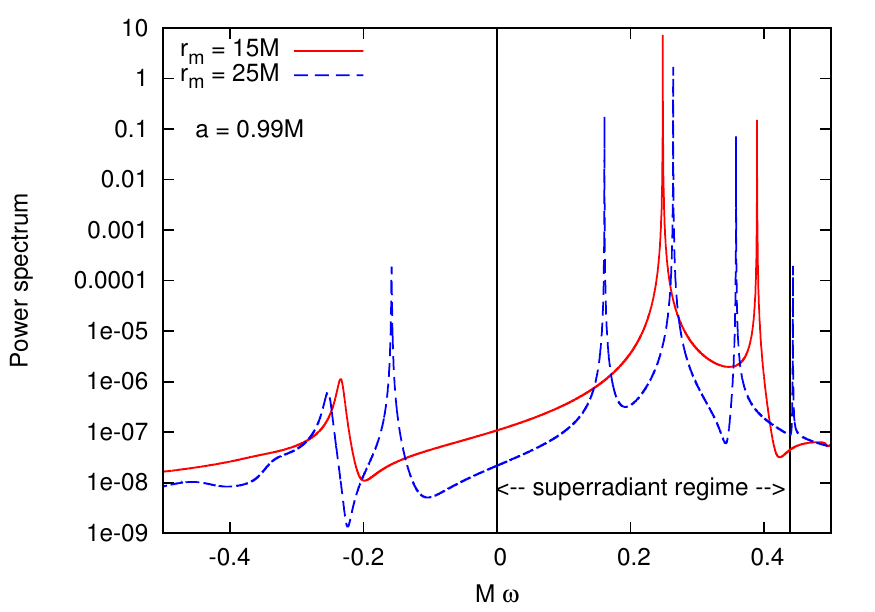}
 \caption{Power spectrum of the scalar field on BH with mirror. The plot shows the power spectrum (\ref{eq:power}) as a function of frequency $M\omega$, computed via a Fourier transform of the time-domain data (Fig.~\ref{fig:timeseries2}), for two mirror radii, $r_m = 15M$ and $r_m = 25M$. Modes inside the superradiant regime, $0 < \omega < m\Omega$, are subject to exponential growth. The modes are approximately evenly-spaced in frequency [Eq.~\ref{eq:mirrorspec}]. The plot illustrates that the effect of increasing the mirror radius $r_m$ is to bring more overtones within the superradiant window.}
 \label{fig:power-spec-mirror}
\end{figure*}

Figure \ref{fig:power-spec-mirror} shows typical power spectra, for mirror radii $r_m = 15M$ and $r_m = 25M$. The sharp peaks indicate the presence of modes in the spectrum. The plot makes it clear that the modes within the superradiant regime $0 < M \omega < m \Omega$ have the largest amplitude (highest peaks), as expected. 

The spectrum has all the features anticipated in the frequency-domain study of the mirrored case, Ref.~\cite{Cardoso:2004-bomb}. The peaks are approximately evenly-spaced in frequency, with a separation that decreases as the mirror radius increases, as predicted by Eq.~(\ref{eq:mirrorspec}). 

The width of a peak in the power spectrum provides information on the decay/growth rate of the corresponding mode.  Consider a monochromatic mode with time dependence $\exp\left(- i (\hat{\omega} t  + i \nu) t \right)$. Its contribution to the power spectrum is
\beq
P(\omega) \approx 1 / \left[ (\omega - \hat{\omega})^2 + \nu^2 \right].
\eeq
Hence, fitting a quadratic to $1/P(\omega)$ in the vicinity of a minimum $\hat{\omega}$ should give an estimate of the decay/growth rate $\nu$, whose sign may be found by examining $f(\omega)$). Despite its promise, we found that this method was not sufficiently accurate to numerically estimate the rates of long-lived modes, and so we pursue an alternative method below.

\subsubsection{Growth rates from frequency-filter\label{subsec:filter}}
To study the individual modes in the time-series data, we devised a simple `frequency-filter' method.  First, we identified the modes in the power-spectrum (e.g.~Fig.~\ref{fig:power-spec-mirror}) using a peak-finding algorithm, and used interpolation to obtain a better estimate for the real frequencies $\hat{\omega}_{n}$.  Next, for each mode, we (i) multiplied the Fourier transform by a narrow filter function centred on the peak, i.e. $f(\omega) \rightarrow f(\omega)F(\omega - \hat{\omega}_{n})$ with
\beq
F(x) = \exp( - x^4 / w^4 ) 
\eeq
and $w = k \Delta \omega$ with $n \sim 5$ (typically) and $\Delta \omega$ defined above Eq.~(\ref{eq:power}); (ii) reconstructed the modal contribution by taking the inverse Fourier transform of the filtered data; (iii) computed the square magnitude of the reconstructed field as a function of time; and (iv) computed the growth index using linear regression on the logarithm of the field magnitude.

\begin{figure}
 \includegraphics[width=8cm]{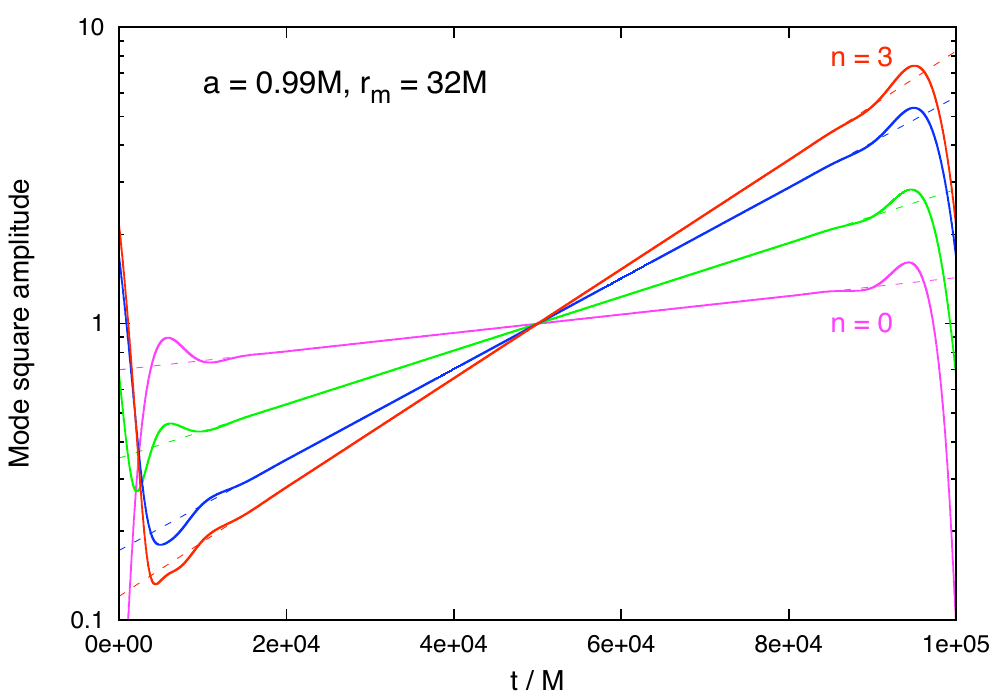}
 \caption{The frequency-filter method. Each solid line in the plot corresponds to the (normalised) square amplitude of a particular mode extracted from the time-domain data. Each mode is isolated by applying a narrow frequency filter to the Fourier transform of the time domain signal, and then applying the inverse transformation (see text). The dotted lines represent the best fits. The plot shows the overtones $n = 0$ to $3$ for $r_m = 32M$ and $a = 0.99M$, for which case $n=3$ is the most rapidly-growing.}
 \label{fig:freqfilter}
\end{figure}

These steps are illustrated in Fig.~\ref{fig:freqfilter}, which shows the square magnitude of the reconstructed field modes for the first few overtones. It shows that the reconstructed field modes at early and late times depend somewhat on the shape of the filter; but that the field around $t_{\text{max}}/2$ is a faithful reconstruction of the modal contribution. 

\begin{figure*}
 \includegraphics[width=8.4cm]{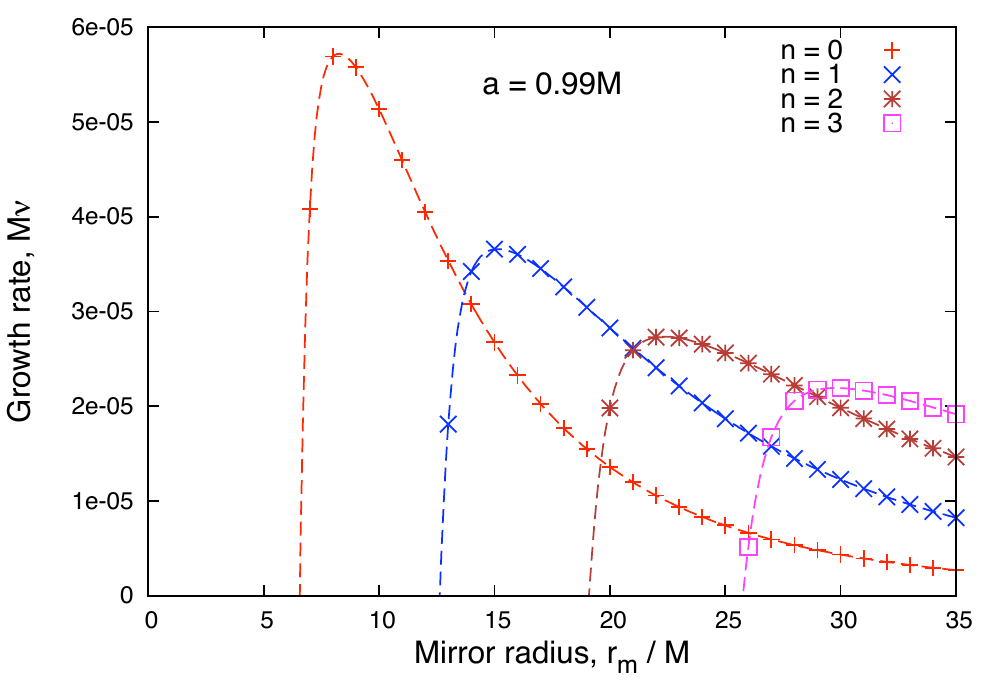}
 \includegraphics[width=8.4cm]{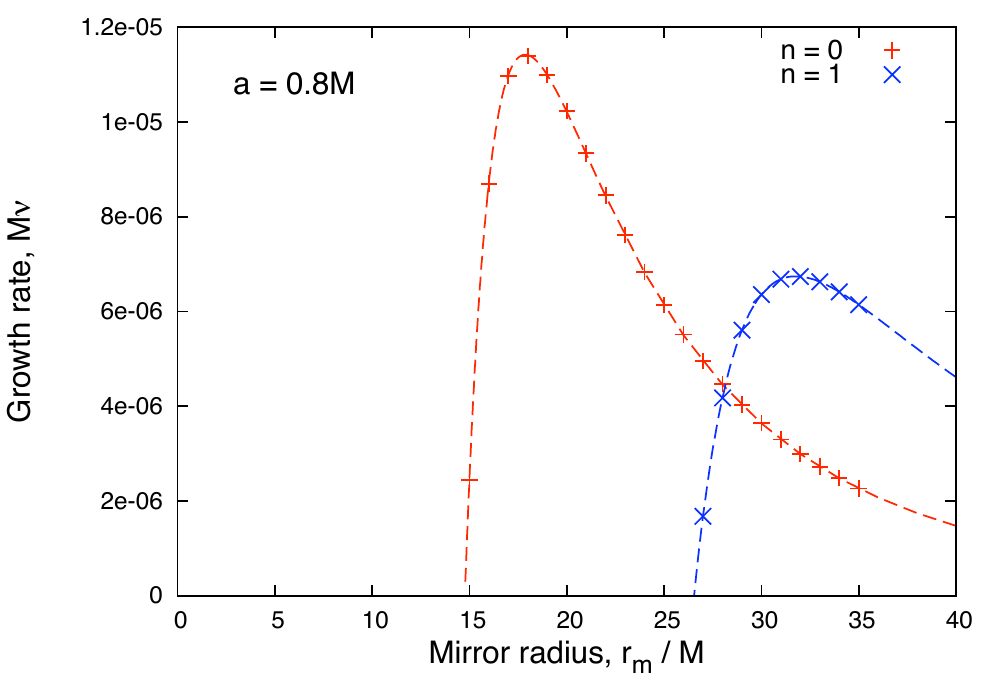}
 \caption{The growth rate of the scalar field for rotating BH with mirror. The plot shows the growth rate of the first few exponentially-growing modes ($\propto \exp(\nu t)$), as a function of mirror radius $r_m$. The points show the growth rates calculated from time-domain data, using runs up to $t = 10^5M$, by applying the frequency-filter method (Sec.~\ref{subsec:filter}). The lines show the growth rates found in the frequency-domain analysis in Ref.~\cite{Cardoso:2004-bomb} (with thanks to the authors).}
 \label{fig:mirror-rates}
\end{figure*}

With the frequency-filter method, we may obtain accurate estimates for the frequency and growth/decay rate of the subdominant modes that are `hidden' in the time-domain data. Figure \ref{fig:mirror-rates} shows estimates of the growth rate of the instability, as a function of mirror radius, for the first few overtones. These estimates are compared with the growth rates obtained in Ref.~\cite{Cardoso:2004-bomb} via a frequency-domain analysis. The agreement is found to be excellent. It is gratifying that many overtones can be extracted from the time domain data, even in cases when the late-time signal is dominated by the fastest-growing mode. 

\subsection{Massive scalar field\label{subsec:massive}}
In this section, we consider the case of a massive scalar field in vacuum. Non-zero mass leads to a spectrum of quasi-bound states with $\text{Re}(\omega) \lesssim \mu$, as summarized in Sec.~\ref{subsec:bound-states}. 

\begin{figure*}
 \includegraphics[width=11cm]{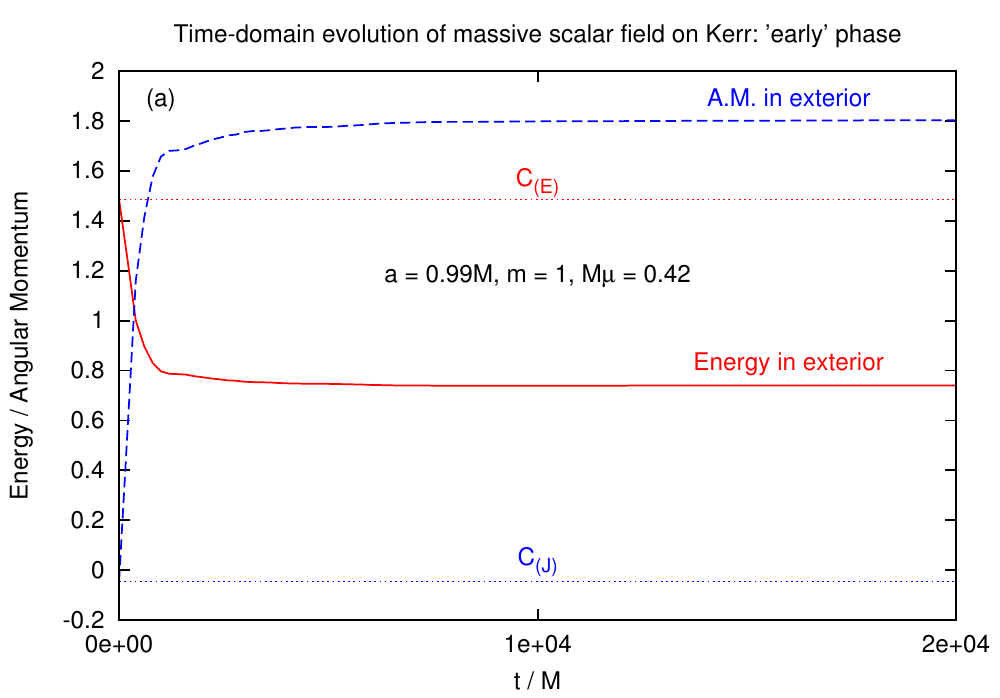} \\
 \includegraphics[width=11cm]{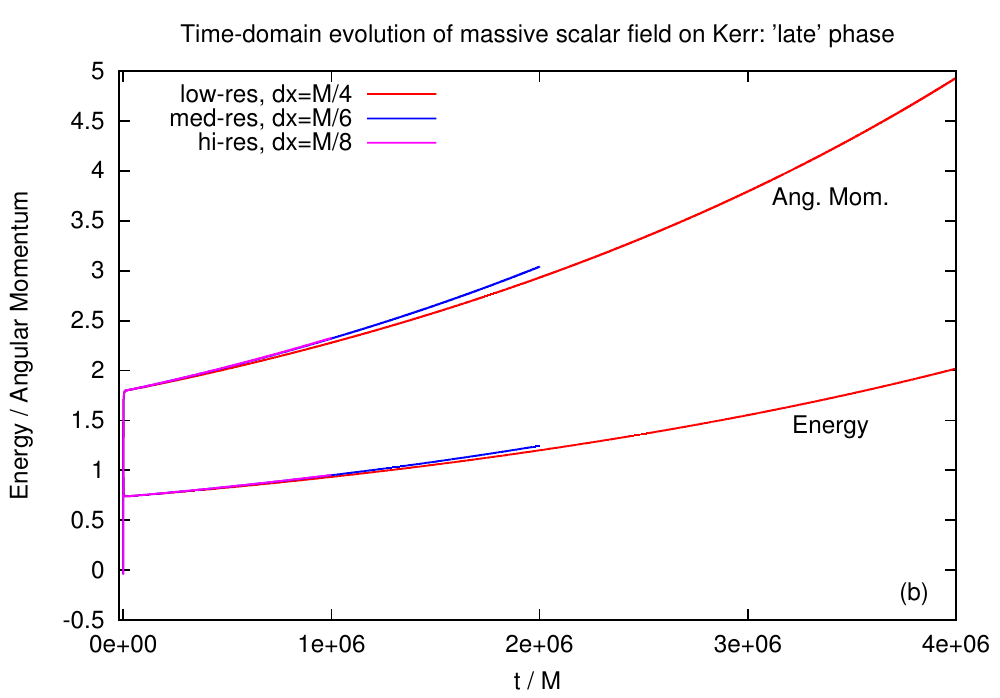} \\
 \includegraphics[width=11cm]{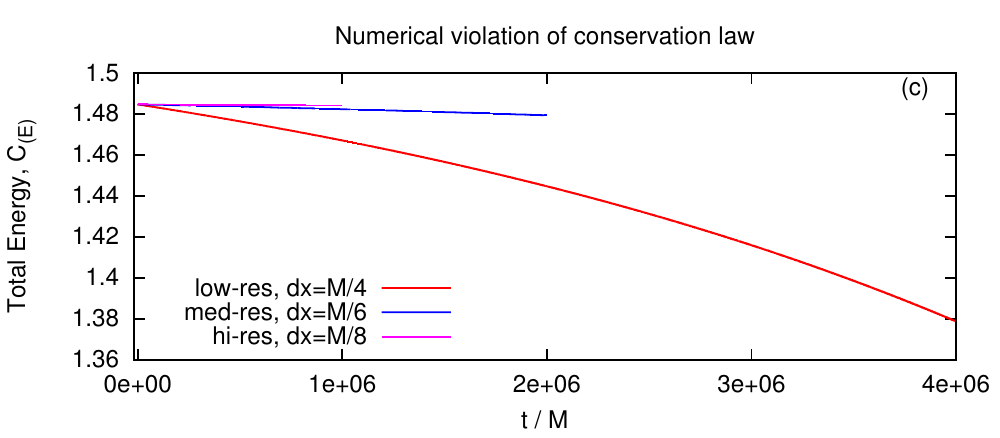}
 \caption{
  Time-domain evolution of a massive scalar field ($M \mu = 0.42$) on rapidly-rotating BH spacetime ($a = 0.99M$) with time-symmetric Gaussian initial data, with $\xmid = 20M$ and $\xwid = 10M$ [see Eq.~(\ref{eq:gaussian})]. The plots show the evolution of the energy $E$ and azimuthal angular momentum (AAM) $J_z$ in the exterior spacetime (\ref{eq:EJ}), and the `conserved' quantities (\ref{SE-constants}). In the initial phase, $t  \lesssim 10^4 M$ shown in plot (a), $E$ decreases and $J_z$ increases as non-superradiant flux, particularly the counter-rotating part of the field with $\omega < 0$, is absorbed by the BH. The `total' energy and AAM, $\CE$ and $\CJ$ [Eq.~(\ref{SE-constants})], are numerically conserved to a good approximation. In the `late' phase, $t \gtrsim \times 10^{4} M$ shown in plot (b), the exterior energy and AAM increase, as growth in the superradiant quasi-bound states begins to dominate. Plot (c) shows that there is some numerical violation of the constraints which diminishes as resolution improves, and which is consistent with the resolution-dependence of the curves at late times seen in plot (b).
 }
 \label{fig:psi-mu}
\end{figure*}

Figure \ref{fig:psi-mu} shows the evolution of the energy and angular momentum of the massive scalar field in the exterior spacetime, for a typical simulation of the $m=1$ mode of a scalar field with $M \mu = 0.42$ and Gaussian initial data, on a Kerr BH with $a = 0.99M$. At `early' times $t \lesssim 10^4 M$, which have been probed by other studies \cite{Strafuss:2005, Yoshino:Kodama, Witek}, we see that the energy in the exterior initially drops off, as non-superradiant flux is absorbed by the BH, before approaching an apparent `stationary' equilibrium. However, `late' time evolutions $t \gg 10^4 M$, reveal that the solution is not truly stationary, as the slow exponential growth in the quasi-bound states begins to dominate. 

To check the validity of our results at late times, we have computed the energy and AM `constants', defined in Eq.~(\ref{SE-constants}), from the numerical data. Figure \ref{fig:psi-mu}(c) shows that there is a slow `drift' in the numerical values of the constants, but that the drift is strongly resolution-dependent, as expected. We also see that the growth rate is resolution-dependent, but that it converges to a physical value as $n$ is increased. 

\subsubsection{Quasi-bound state spectrum}

\begin{figure*}
 \includegraphics[width=11cm]{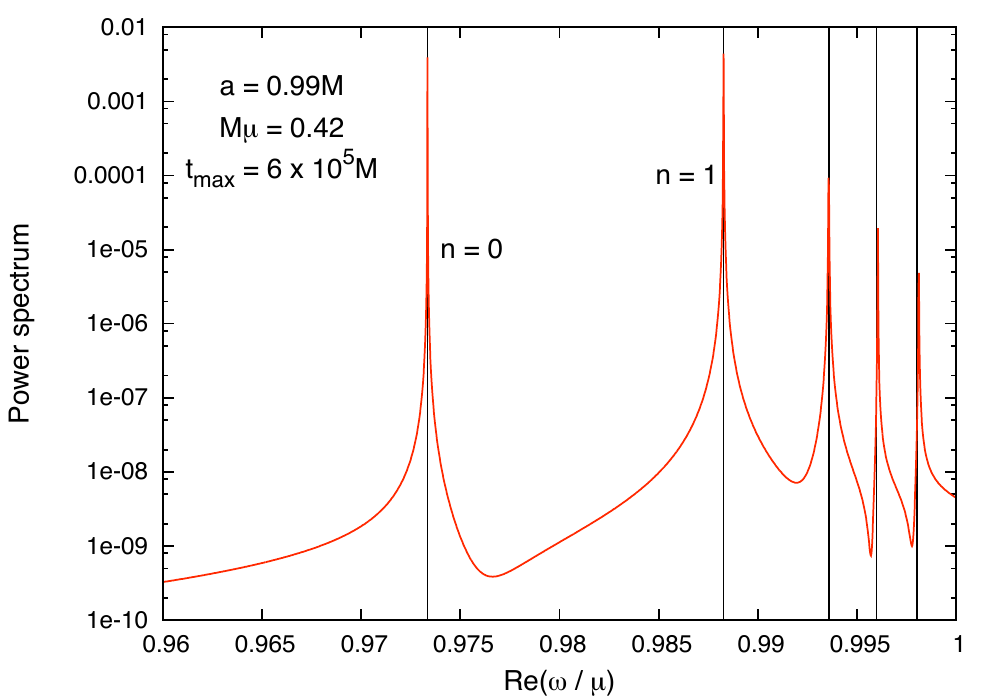}
 \caption{Power spectrum of massive scalar field ($M\mu = 0.42$) on fast-rotating BH spacetime ($a = 0.99M$) in the co-rotating dipole $l=m=1$. The spectrum is computed from time-domain response to Gaussian initial data, up to $t = 6 \times 10^5 M$. The fundamental mode and the first four overtones are visible. The frequencies of the first quasi-bound states, $n=0\ldots 4$ for $l = m = 1$, computed by frequency-domain analysis, are shown at $\omega / \mu \approx 0.9733$, $0.9883$, $0.9936$, $0.9960$ and $0.9980$. }
 \label{fig:power-spec-mu}
\end{figure*}

By extracting the field at a fixed point $x=x_1$, and taking the Fourier transform, we may obtain the power spectrum $P_l(\omega)$ of the massive field (see Sec.~\ref{subsec:mirrorspec}). Figure \ref{fig:power-spec-mu} shows a typical spectrum in the case $a = 0.99M$, $M \mu = 0.42$. The fundamental mode and the first four overtones, $n = 0 \ldots 4$, are apparent as sharp peaks in the spectrum. With ultra-long timescale simulations, we gain sufficient frequency resolution to determine the bound-state frequencies with high accuracy. Figure \ref{fig:power-spec-mu} shows a comparison between the frequencies found by Fourier analysis of the time domain signal, and the frequencies found from frequency-domain analyses (e.g.~Sec.~\ref{subsec:bound-states} and Ref.~\cite{Dolan:2007}).

\begin{figure*}
 \includegraphics[width=11cm]{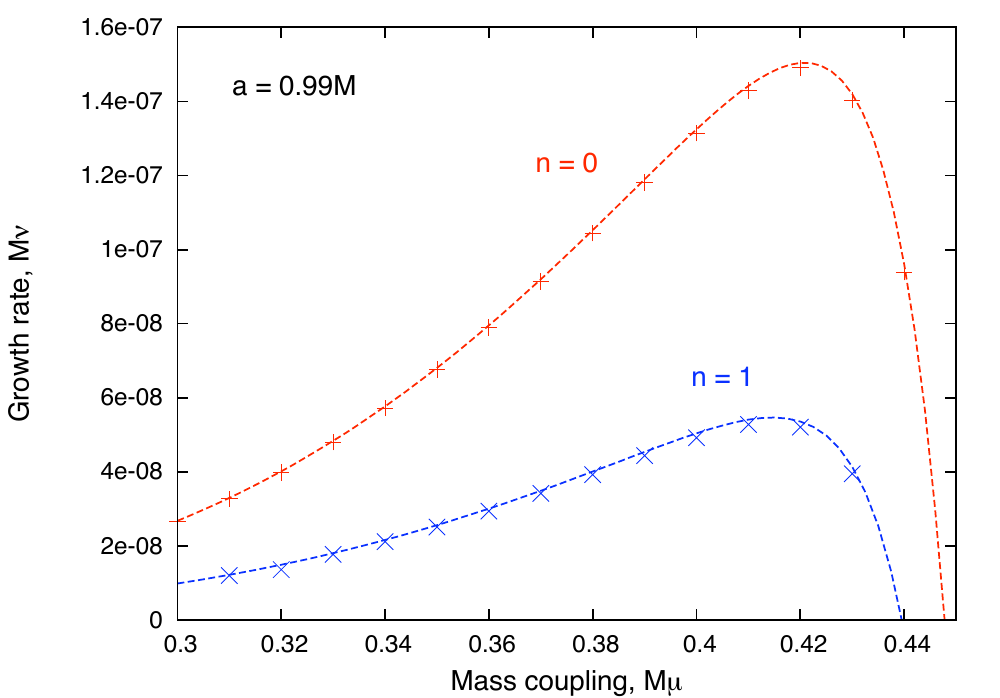}
 \caption{Instability growth rates of massive scalar field on rapidly-rotating Kerr spacetime ($a = 0.99M$). 
 The lines show the growth rates of the fundamental and first-overtone bound states of the co-rotating dipole $l=m=1$, established via the frequency-domain method \cite{Dolan:2007}. The points show estimates of the growth rates found by applying the frequency-filter method (Sec.~\ref{subsec:filter}) to data extracted from time-domain runs (with $t_{\text{max}} = 10^5M$ and resolution $dx = M / 6$). The observed agreement implies that the quasi-bound state spectrum can be accurately determined from `ultra-long' time-domain datasets in cases, such as the Proca field on Kerr spacetime, where a lack of separability impedes a frequency-domain analysis. 
 }
 \label{fig:mu-rates}
\end{figure*}

By applying the frequency-filter method described in Sec.~\ref{subsec:mirrorspec} we may estimate the growth rate of the superradiant quasi-bound states. A key advantage of the method is that it allows us to separately analyse the many overtones that are in the spectrum, even though the time-domain signal may be dominated by a single mode. Furthermore, we find that we may extract good estimates for the growth rate of lowest overtones $n= 0, 1$ from runs of $t_\text{max} = 10^5 M$ which are relatively `short' in comparison to the e-folding time ($\nu^{-1} > 5.8 \times 10^6 M$) of the instability. Figure \ref{fig:mu-rates} shows some typical results, comparing the growth rate extracted from time-domain simulations, with the much more accurate values found by frequency-domain analysis \cite{Dolan:2007}. The agreement is good, reinforcing once more the concordant view of quasi-bound states described in Sec.~\ref{subsec:bound-states}. 

\section{Discussion and conclusion\label{sec:conclusion}}
In this paper we have described a `coupled 1+1D' time-domain scheme for evolving the scalar field on Kerr spacetime. The scheme enabled us to study the evolution of the scalar field, within the linear regime, over ultra-long timescales $t \sim 10^6M$, which represents a factor of 100 to 1000 improvement over previous time-domain studies \cite{Strafuss:2005, Yoshino:Kodama, Witek}. The efficiency of the scheme allowed us to track the massive-field superradiant instability for a significant fraction of an e-folding time. We have shown that accurate values for the quasi-bound state frequencies and growth rates may be extracted directly from time-domain data via Fourier analysis. We found that these values were in excellent agreement with the results of frequency-domain analyses (see Fig.~\ref{fig:mirror-rates} and Fig.~\ref{fig:mu-rates}). Thus, our technique offers a promising way forward for tackling scenarios where a frequency-domain analysis, reliant on the separability of the underlying equations, is not feasible. The method would seem to be ripe for application to the Proca-Kerr system \cite{Rosa:Dolan, Pani:PRD, Pani:PRL}, recently examined by Witek \emph{et al.} \cite{Witek} in the time domain for the first time.

The efficiency of the method is primarily due to the dimensional reduction procedure, i.e., working in a 1+1D rather than 3+1D domain. In other words, the key step in our method is the decomposition of the field into a superposition of harmonics. Even though this does not lead on to a \emph{full} decoupling, it turns out that the coupling between $l$ modes is very simple. In the scalar-field case, the coupling is only between the nearest-neighbour modes of the same parity. In the Teukolsky equation ($s \neq 0$), we anticipate the coupling would also be simple, but instead between nearest- and next-to-nearest neighbours \cite{Hughes:2000}. In the higher-spin massive-field case (e.g.~Proca) the coupling may be substantially more complicated, although the slow-rotation expansion of \cite{Pani:PRD, Pani:PRL} suggests it may be tractable.

Our 1+1D scheme is similar in some regards to that developed in Ref.~\cite{Racz:Toth, Csizmadia}, to look at the late-time tails of massless fields on Kerr spacetime. In Ref.~\cite{Racz:Toth}, the coefficient of $\partial_{tt} \Psi$, which depends on $\cos^2 \theta$, appears in the denominator on the right-hand side of the equation. The denominator is converted to an infinite series in the numerator, which leads to coupling between all $l$ modes after harmonic decomposition (see Sec.~2.2 in \cite{Racz:Toth}). Fortunately the couplings die away rapidly with increasing $l$ and the method is useful in practice. An advantage of the approach of Ref.~\cite{Racz:Toth} is that a matrix-inversion step (Sec.~\ref{subsec:FD}) is not required at each time step. This may offer an alternative route to tackling the Proca-Kerr case, where coupling terms may be substantially more complicated.

Our method also makes use of a technique from computational electrodynamics: the Perfectly Matched Layer \cite{Berenger:1994, Taflove:Hagness}. We have shown that the PML offers a numerically-robust way of implementing the absorbing boundary condition at the event horizon.  It seems that this technique may be readily incorporated into a range of linear wave equations, and yet to date it not been used widely in black hole perturbation theory. The PML also offers distinct advantages in multi-dimensional scenarios, where stable boundary conditions are more difficult to implement. It would be interesting to compare the efficacy of the PML with alternative methods, such as horizon-penetrating coordinate systems with excision. 

We have run our time-domain code for a range of mass couplings, $0.1 \lesssim M \mu \lesssim 1$, and BH spin rates $0 \le a/M \le 0.99$, for low azimuthal numbers $m$. In every case, we have found that the growth of the field (if present) is due to growth in a superposition of quasi-bound states which lie within the superradiant regime. Furthermore, these quasi-bound states are found to have the expected frequencies and growth rates from the `concordant' frequency-domain analyses (see Sec.~\ref{subsec:bound-states}). We have found no evidence in favour of the more rapid growth rates reported in \cite{Strafuss:2005} and \cite{Hod:Hod:2009a}.  

Unabated, an exponentially-growing instability will inflate the magnitude of the scalar field until non-linear effects can no longer be ignored. Non-linear effects arise from (at least) two sources: the self-interaction of the field, and the back-reaction of the field upon the spacetime. In the context of axions, the former effect may be modelled by replacing $\mu^2 \Phi$ with $\mu^2 \sin \Phi$ in the scalar-field equation, introducing a non-linear term at leading order $\mu^2 \Phi^3$. A recent pioneering time-domain study \cite{Yoshino:Kodama} looked at the effect of non-linear coupling. It concluded that, whereas small self-couplings will increase the rate of energy extraction, large self-couplings will cause a collapse of the axion cloud. This suggests that the endpoint of the instability is an explosive phenomenon \cite{Arvanitaki1}. As yet, the transition from small to large couplings has not been tracked via a time-domain simulation, due to strong limitations on run-time imposed by the computational expense of 3+1D codes. It may be possible that a dimensionally-reduced code could allow longer run-times; however, the non-linear terms would introduce couplings between $m$ and $l$ modes, which would substantially increase the complexity of implementation. A perhaps more tractable problem would be to use a time-domain code to study the generation of gravitational waves (i.e.~linearized metric perturbations on the Kerr background) induced by the stress-energy of the scalar field.

A further prospect is to apply dimensional reduction to study the Proca field over long timescales within the linear regime. Here a $2+1$D code, exploiting a simple separation into $m$-modes, would seem suitable. A $2+1$D code would allow one to probe beyond the timescales achieved in Ref.~\cite{Witek} ($t_{\text{max}} = 3 \times 10^3M$). The resolution of the power spectrum scales with run-time, and so, with long-timescale datasets, one would be able to isolate the modes by applying a narrow filter to peaks in the power spectrum (as in Sec.~\ref{subsec:filter}) to determine the details of the quasi-bound states (i.e.~frequencies and growth rates) to high precision. This is surely an interesting challenge. 

In summary, time-domain simulations offer a way to examine the onset, development and termination of black holes superradiant instabilities which may be ignited in nature by ultra-light bosonic fields. It is important to understand these instabilities fully, if we are to use new observations of astrophysical black hole parameters to constrain the boson mass spectrum \cite{Arvanitaki1,Arvanitaki2,Kodama:Yoshino,Yoshino:Kodama,Pani:PRL,Pani:PRD}. Furthermore, the study of field configurations which are long-lived or quasi-stable is of foundational interest, because such configurations would seem to challenge to the spirit of the `no-hair' conjecture. We have shown here that the scalar-field instability has been well understood within the linear regime. There remains much scope for pursuing the instability into the non-linear regime, and in the context of massive vector fields.

\begin{acknowledgments}
I am grateful to Joao Rosa, Helvi Witek, Barry Wardell, Elizabeth Winstanley and Leor Barack for helpful discussions, and to Vitor Cardoso and Shijun Yoshida for providing data for validation and comparison. I acknowledge the use of the following HPC resources: Iridis at the University of Southampton, Tesla at University College Dublin, and Iceberg at the University of Sheffield. Financial support in the early stages of this work was provided by the Engineering and Physical Sciences Research Council (EPSRC) under Grant No. EP/G049092/1.
\end{acknowledgments}

\bibliographystyle{acm}

\end{document}